\documentclass[
reprint,
longbibliography,
amsmath, 
amssymb, 
aps, 
superscriptaddress,
prx,
]
{revtex4-2}
\usepackage{graphicx}
\usepackage{dcolumn}
\usepackage{bm}
\usepackage{braket} 
\usepackage{hyperref}
\usepackage{subfigure} 
\usepackage{nccmath}
\usepackage{amsthm} 
\usepackage{mathrsfs} 
\usepackage{xfrac} 
\usepackage{pifont}
\usepackage[export]{adjustbox}
\usepackage{bbm} 
\usepackage[utf8]{inputenc} 
\usepackage[normalem]{ulem}
\usepackage[dvipsnames,svgnames,table]{xcolor}
\usepackage[sort&compress]{natbib}
\usepackage{color}
\usepackage{longtable}
 
\usepackage{commath}
\usepackage{multirow}
\usepackage{array}

\begin{document}

\title{Evaluating the performance of quantum processing units at large width and depth}

\author{J. A. Monta\~nez-Barrera}
    \altaffiliation{j.montanez-barrera@fz-juelich.de}
	\affiliation{Jülich Supercomputing Centre, Institute for Advanced Simulation, Forschungszentrum Jülich, 52425 Jülich, Germany}
    
\author{Kristel Michielsen}
	\affiliation{Jülich Supercomputing Centre, Institute for Advanced Simulation, Forschungszentrum Jülich, 52425 Jülich, Germany}
	\affiliation{AIDAS, 52425 Jülich, Germany}
	\affiliation{RWTH Aachen University, 52056 Aachen, Germany}
 \author{David E. {Bernal Neira}}
    \altaffiliation{dbernaln@purdue.edu}
	\affiliation{Davidson School of Chemical Engineering, Purdue University, 47907, West Lafayette, Indiana, USA}

\begin{abstract}
Quantum computers have now surpassed classical simulation limits, yet noise continues to limit their practical utility.  
As the field shifts from proof-of-principle demonstrations to early deployments, there is no standard method for meaningfully and scalably comparing heterogeneous quantum hardware.  
Existing benchmarks typically focus on gate-level fidelity or constant-depth circuits, offering limited insight into algorithmic performance at depth.  
Here we introduce a benchmarking protocol based on the linear ramp quantum approximate optimization algorithm (LR-QAOA), a fixed-parameter, deterministic variant of QAOA.  
LR-QAOA quantifies a QPU’s ability to preserve a coherent signal as circuit depth increases, identifying when performance becomes statistically indistinguishable from random sampling.  
We apply this protocol to 24 quantum processors from six vendors, testing problems with up to 156 qubits and 10,000 layers across 1D-chains, native layouts, and fully connected topologies. This constitutes the most extensive cross-platform quantum benchmarking effort to date, with circuits reaching a million two-qubit gates. LR-QAOA offers a scalable, unified benchmark across platforms and architectures, making it a tool for tracking performance in quantum computing.

\begin{description}
	\vspace{0.2cm}
	\item[Keywords] Quantum Benchmarking, IBM Heron, IBM Eagle, IQM Garnet, Quantinuum H2-1, Rigetti Ankaa-3, IonQ Aria, Forte, OriginQ Wukong, LR-QAOA, Weighted MaxCut.
\end{description}

\end{abstract}

\maketitle
\section{Introduction}\label{Sec:Introduction}

\begin{figure*}[t]
\centering
\includegraphics[width=18cm]{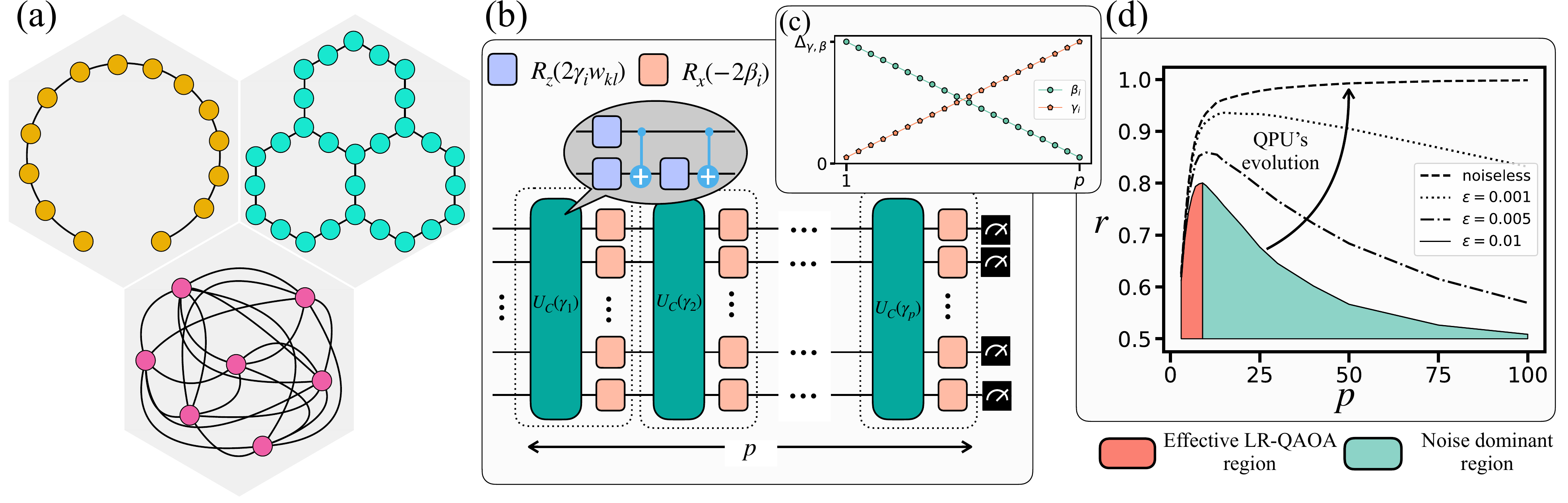}
\caption{\label{Fig:Scheme}Scheme for benchmarking quantum processing units (QPUs) using LR-QAOA.
(a) Graph topologies used in the benchmark: 1D-chain (yellow), native layout (NL, green), and fully connected (FC, pink).
(b) QAOA circuit structure consisting of alternating layers of the problem and mixer Hamiltonians; $p$ denotes the circuit depth.
(c) Linear parameter schedule in LR-QAOA; the slope is given by $\Delta_{\gamma,\beta}/p$.
(d) Expected behavior of the approximation ratio as a function of depth $p$ under varying levels of depolarizing noise \(\varepsilon\), shown as black curves.}
\end{figure*}

In the current stage of quantum technology, quantum processors have reached sizes that exceed exact classical simulation capabilities.
However, extracting a reliable computational value from these devices remains limited by noise.
As hardware scales in both qubit count and circuit depth, benchmarking tools must evolve to reflect meaningful progress toward practical quantum advantage.

A central challenge is comparing heterogeneous quantum devices in a scalable and interpretable way.
Existing metrics such as quantum volume (QV) and error-per-layered-gate (EPLG) offer partial views, focusing on circuit reach or gate fidelity, but fail to capture algorithmic performance under realistic workloads.
There is a growing need for platform-agnostic benchmarks that assess usable circuit depth, coherence retention, and whole-system behavior across architectures.

We address this gap with the linear ramp quantum approximate optimization algorithm (LR-QAOA)~\cite{Montanez-Barrera2024b}, a fixed-parameter variant of QAOA in which depth increases linearly and parameters evolve deterministically.
Its non-variational schedule avoids optimization overhead, and its layered structure supports straightforward scaling in both width and depth.
These properties make LR-QAOA a practical, architecture-neutral benchmark for evaluating QPU performance at scale.

The key performance metric is the approximation ratio $r$, which grows with depth in the absence of noise and degrades as coherence is lost.
This transition reveals the usable depth of a device, and we demonstrate it using LR-QAOA circuits with up to 156 qubits and 10{,}000 layers.

This paper introduces a volumetric benchmark based on LR-QAOA applied to Weighted MaxCut (WMC) instances across three graph topologies: 1D chain, native layout, and fully connected. We benchmark 24 QPUs from six vendors, IBM, IQM, IonQ, Quantinuum, Rigetti, and OriginQ, covering problem sizes from \(N_q=5\) to \(N_q=156\) and circuit depths up to \(p=10,000\). Our results reveal depth regimes where noise overtakes the LR-QAOA signal, with thresholds varying across hardware architectures and generations. We also demonstrate that LR-QAOA enables consistent cross-platform comparisons and uncovers performance trends not captured by existing metrics such as QV or EPLG.

The remainder of the paper is organized as follows.
Section~\ref{Sec:Methods} describes benchmarking methodologies for QPUs, LR-QAOA, and the experimental setup. In Sec.~\ref{Sec:Results}, we present the results for the graph benchmarks. Finally, Sec.~\ref{Sec:Conclusions} summarizes the main conclusions.

\section{Methods}\label{Sec:Methods}

This section presents the LR-QAOA protocol, which serves both as a benchmarking method for quantum processing units (QPUs) and as an algorithm capable of approximating solutions to combinatorial optimization problems.
We begin by describing the LR-QAOA framework, including its problem encoding, circuit structure, and performance metric.
We then outline the experimental setup used to implement LR-QAOA on the QPUs.
Finally, we provide context by comparing LR-QAOA to existing quantum benchmarking approaches.

\subsection{LR-QAOA}

The linear ramp quantum approximate optimization algorithm (LR-QAOA) is a fixed-parameter, non-variational variant of QAOA~\cite{Farhi2014}, with linearly ramped parameters, producing circuits that are deterministic, scalable, and easy to implement across platforms.
As the depth increases, the algorithm typically improves solution quality until noise begins to dominate, making it well-suited to probe the useful depth limits of quantum hardware.
LR-QAOA serves both as a benchmarking tool and an approximate solver for combinatorial optimization problems, with performance evaluated via classical post-processing.

Figure~\ref{Fig:Scheme} summarizes the LR-QAOA benchmarking setup.
Figure~\ref{Fig:Scheme}(a) shows the three graph topologies used in our tests: 1D-chain, native layout (NL), and fully connected (FC).
The first two are suitable for fixed-layout devices, e.g., IBM's heavy-hex architecture~\cite{IBMHeavyHexLattice}, while the FC benchmark is designed for both fixed-layout and fully connected architectures, such as \texttt{quantinuum\_H2-1}.
Figure~\ref{Fig:Scheme}(b) depicts the QAOA circuit structure used in the benchmark. It consists of alternating applications of problem and mixer unitaries over $p$ layers.
The parameters used in these unitaries follow a linear ramp schedule, as shown in Fig.~\ref{Fig:Scheme}(c), which approximates adiabatic evolution via first-order Trotterization~\cite{farhi2000quantum}.
Finally, Fig.~\ref{Fig:Scheme}(d) illustrates the typical behavior of the approximation ratio $r$ as a function of depth $p$ under different noise levels.
The red region corresponds to algorithm-dominated performance, and the green region to noise-dominated behavior.
As devices improve, we expect the crossover between these regions to occur at larger $p$.

We benchmark quantum devices by solving instances of the WMC problem using LR-QAOA.
WMC is defined on a weighted graph $\mathcal{G} = (V, E)$ and aims to partition its vertices $V$, into two groups such that the sum of the edge weights between the groups is maximized.
This problem is NP-hard~\cite{karp1975reducibility} and equivalent to finding the ground state of an Ising spin glass model defined on the same graph~\cite{barahona1982computational}.
It is well-suited for benchmarking due to its generality, two-qubit structure, and efficiently computable optimal value.

The WMC cost Hamiltonian is given by
\begin{equation}
H_C = \sum_{\{i,j\} \in E(\mathcal{G})} w_{ij} \sigma^z_i \sigma^z_j,
\end{equation}
where $\sigma^z_i$ is the Pauli-$Z$ operator on qubit $i$, and $w_{ij}$ is the weight of the edge between vertices $i$ and $j$. Each qubit represents a node in the graph.
The problem Hamiltonian is encoded as a unitary gate
\begin{equation}
U_C(H_C, \gamma_k) = e^{-i \gamma_k H_C},
\end{equation}
where $\gamma_k$ is taken from the linear ramp schedule.

The mixer Hamiltonian is defined as
\begin{equation}
H_B = \sum_{i=0}^{N_q - 1} \sigma^x_i,
\end{equation}
and the corresponding mixer unitary is
\begin{equation}
U_B(H_B, \beta_k) = e^{-i \beta_k H_B},
\end{equation}
where $\sigma^x_i$ is the Pauli-$X$ operator on qubit $i$, and $\beta_k$ is the ramped parameter at layer $k$.

The ramp schedules for $\beta_k$ and $\gamma_k$ are defined as
\begin{equation}
\beta_k = \left(1 - \frac{k}{p}\right) \Delta_\beta, \qquad
\gamma_k = \frac{k+1}{p} \Delta_\gamma,
\end{equation}
for $k = 0, \dots, p-1$.
In all experiments, we set $\Delta_\beta = \Delta_\gamma = \Delta_{\beta,\gamma}$.
We choose well-conditioned ramp schedules with observable signal growth in noise-free settings and balanced saturation under noise, validated in prior simulations~\cite{Montanez-Barrera2024b}. The specific values used are listed in Sec.~\ref{A:Deltas}.

The circuit is initialized in the uniform superposition state $\ket{+}^{\otimes N_q}$.
We execute the full LR-QAOA circuit of depth $p$, sample $n$ bitstrings $x_1, \dots, x_n$, and compute the approximation ratio
\begin{equation}
r = \frac{\frac{1}{n} \sum_{i=1}^n C(x_i)}{C(x^*)},
\label{eq:r}
\end{equation}
where $x^*$ is the optimal bitstring and $C(x)$ is the WMC cost function given by
\begin{equation}
C(x) = \sum_{\{k,l\} \in E(\mathcal{G})} w_{kl}(x_k + x_l - 2x_k x_l).
\end{equation}
The optimum $x^*$ is computed using classical solvers such as CPLEX~\cite{Bliek2014SolvingMQ}.
The approximation ratio $r$ increases with circuit depth $p$ until it saturates or decays due to noise, making it a useful indicator of the effective computational depth a QPU can sustain.

\subsection{Experimental Setup}

Table~\ref{Tab:properties} summarizes the number of qubits \(N_q\), two-qubit gate duration, average two-qubit gate error, and native layout connectivity for a subset of the QPUs used in this work.
Device specifications were obtained from vendor documentation: IBM devices from~\cite{ibmquantumservices}; IonQ QPUs from~\cite{microsoftionqaria}; \texttt{quantinuum\_H1-1} and \texttt{quantinuum\_H2-1} from~\cite{quantinuumh1,quantinuumh2}; \texttt{iqm\_garnet} from~\cite{IQM2024Garnet}; Rigetti Ankaa from~\cite{Rigetti}; and \texttt{originq\_wukong} from~\cite{OriginQuantum}.
We could not find publicly available information for the two-qubit gate time of \texttt{quantinuum\_H1-1} and \texttt{originq\_wukong}.

The number of two-qubit gates and depth required to implement $p$ layers of LR-QAOA depends on the device architecture and native gate set.
Gate counts and circuit depths scale linearly with QAOA layer count $p$, with prefactors depending on the QPU’s gate set and connectivity. Detailed formulas are listed in Appendix~\ref{A:circuit_scaling}.

We used 1,000 measurement samples per instance on IBM, IQM, Rigetti, and OriginQ devices; 100 samples on IonQ devices; 50 samples on \texttt{quantinuum\_H2-1} for $N_q \leq 50$; and 7 samples for $N_q = 56$.

IBM recently introduced fractional gates~\cite{ibm_fractional_gates} on Heron devices that directly implement the $ZZ$ and $R_x$ gates required by LR-QAOA.

These gates halve the number of two-qubit operations and reduce the total two-qubit circuit depth by a factor of two compared to CZ-based implementations.
This enables deeper circuits with lower overall noise accumulation and plays a significant role in performance gains seen across Heron-class devices~\cite{ibm_fractional_gates}.

\begin{table}\caption{\label{Tab:properties}Number of qubits ($N_q$), native two-qubit gate duration ($t_{2q}$), average two-qubit gate error ($\varepsilon_{2q}$), and native layout (NL): heavy-hex (HE), square (SQ), or fully connected (FC). }
\begin{tabular}{ c|c|c|c|c|c|}
 \cline{2-5}
 & \multicolumn{4}{c|}{Properties} \\
 \cline{1-5}
 \multicolumn{1}{ |c| }{QPU} & $N_q$ & $t_{2q} $& $\varepsilon_{2q}$& NL\\
 \hline
 \multicolumn{1}{ |c| }{\texttt{ibm\_brisbane}} & 127  & 600 ns & 7.760e-3 & HE\\ 
 \hline

 \multicolumn{1}{ |c| }{\texttt{ibm\_torino}-v1} & 133 & 68/184 ns & 2.977e-3 & HE \\
\hline
 \multicolumn{1}{ |c|  }{\texttt{ibm\_fez}} & 156 & 68 ns & 2.959e-3 & HE \\

 \hline
 \multicolumn{1}{ |c|  }{\texttt{ibm\_marrakesh}} & 156 & 68 ns & 2.157e-3 & HE \\

  \hline
 \multicolumn{1}{ |c|  }{\texttt{ibm\_aachen}} & 156 & 68 ns & 2.396e-3 & HE \\

  \hline
 \multicolumn{1}{ |c|  }{\texttt{ibm\_kingston}} & 156 & 68 ns & 2.954e-3 & HE \\

\hline
 \multicolumn{1}{ |c|  }{\texttt{quantinuum\_H1-1}} & 20 & - & 0.860e-3 & FC \\

 \hline
 \multicolumn{1}{ |c|  }{\texttt{quantinuum\_H2-1}} & 56 & 2000 $\mu$s & 1.300e-3 & FC \\

  \hline
 \multicolumn{1}{ |c|  }{\texttt{ionq\_forte}} & 36 & 970 $\mu$s & 4.000e-3 & FC \\

  \hline
 \multicolumn{1}{ |c|  }{\texttt{ionq\_aria\_2}} & 25 & 600 $\mu$s & 4.000e-3 & FC \\

  \hline
 \multicolumn{1}{ |c|  }{\texttt{ionq\_harmony}} & 11 & 210 $\mu$s & 39.80e-3 & FC \\

   \hline
 \multicolumn{1}{ |c|  }{\texttt{iqm\_spark}} & 5 & $\le$ 100 ns & 10.00e-3 & SQ \\

    \hline
 \multicolumn{1}{ |c|  }{\texttt{iqm\_garnet}} & 20 & 20/40 ns & 5.000e-3 & SQ \\

    \hline
 \multicolumn{1}{ |c|  }{\texttt{rigetti\_ankaa\_2}} & 84 & 68 ns & 55.60e-3 & SQ \\
    \hline
 \multicolumn{1}{ |c|  }{\texttt{rigetti\_ankaa\_3}} & 82 & 72 ns & 59.50e-3 & SQ \\
     \hline
 \multicolumn{1}{ |c|  }{\texttt{originq\_wukong}} & 72 & - & 27.90e-3 & SQ \\
 \hline
\end{tabular}
\end{table}

\subsection{Benchmarking Background}
\label{Sec:BenchmarkingBackground}

Quantum benchmarking protocols aim to characterize the capabilities and limitations of quantum processors in a scalable and interpretable way.  
Among the most widely adopted is \emph{quantum volume} (QV)~\cite{PhysRevA.100.032328}, which measures the largest square-shaped quantum circuit that a device can execute reliably.  
QV has been used by multiple vendors~\cite{IBM2022QuantumVolume, PhysRevX.13.041052, AQT2023QuantumVolume, IQM2024Garnet, quantinuum_hardware_2023}, with the highest reported value to date from \texttt{quantinuum\_H2-1} at $2^{23}$~\cite{quantinuum_qv_2025}.  
However, QV remains limited to small circuit sizes because its validation requires classical simulation, which scales exponentially.  

\emph{Randomized benchmarking} (RB)~\cite{Emerson_2005} offers an alternative that is more scalable but less holistic.  
RB evaluates gate-level fidelity by applying sequences of random Clifford operations and fitting the decay in success probability.  
Variants such as direct RB (DRB)~\cite{Proctor_2019} and mirror RB (MRB)~\cite{Proctor_2022} have been developed to better assess multi-qubit performance.

\emph{Cross-entropy benchmarking} (XEB)~\cite{Boixo2018} has been used primarily in the context of demonstrating quantum supremacy~\cite{Arute2019}, while \emph{algorithmic qubits} (AQ)~\cite{Chen2023} define the number of logical qubits a device can support for a given quantum workload.

Recently, IBM introduced \emph{error per layered gate} (EPLG)~\cite{McKay2023} as a successor to QV.  
EPLG applies randomized benchmarking to disjoint two-qubit subsystems and estimates the average error across them.  
This approach yields localized error rates and scales to large systems, though it does not reflect algorithmic performance directly.

Other proposals focus on \emph{application-oriented} benchmarks~\cite{Lubinski_2023, chen2023benchmarkingtrappedionquantumcomputer, lubinski2024, Miessen2024}, which directly evaluate performance on structured problem instances relevant to optimization or simulation.

While most existing benchmarks focus on gate fidelity or small circuits, our approach targets algorithmic performance degradation as depth increases.

The LR-QAOA benchmark introduced in this work complements these efforts by probing the depth at which algorithmic performance degrades due to noise.
It is scalable, implementation-friendly, and tied to a well-defined optimization metric: the approximation ratio $r$. Because LR-QAOA circuits can be mapped to arbitrary topologies and benchmark problems with known optima, the protocol allows for consistent, cross-platform evaluation of QPUs. Together, these properties position LR-QAOA as a complementary tool to existing protocols, one that connects algorithmic depth to performance limitations.

\section{Results}\label{Sec:Results}

\subsection{1D-chain benchmark}

\begin{figure*}[!tbh]
\centering
\includegraphics[width=18cm]{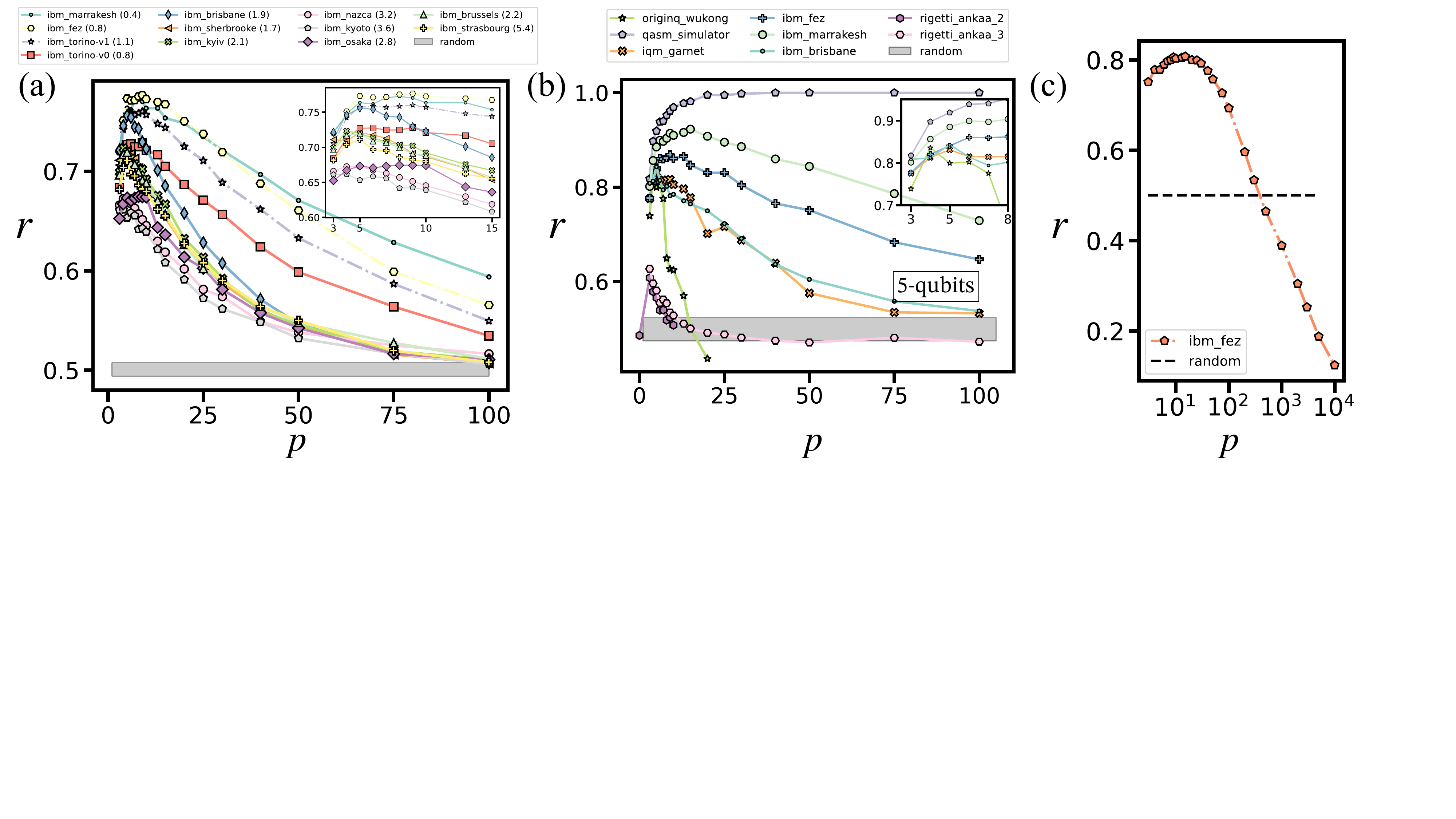}
\caption{\label{Fig:1Dsmall} Approximation ratio versus number of LR-QAOA layers for 1D-chain WMC problems on IBM, IQM, Rigetti, and OriginQ devices.
(a) Performance on a 100-qubit problem across IBM Eagle and Heron QPUs. Numbers in parentheses indicate EPLG at the time of execution.
(b) Cross-platform comparison for a 5-qubit problem run on different QPUs.
(c) Scaling study on \texttt{ibm\_fez} from \(p=3\) to \(p=10{,}000\), reaching 990,000 ZZ gates at the deepest layer using fractional gates.
The dashed line in (c) marks the expected approximation ratio of a random sampler.}
\end{figure*}

The 1D-chain benchmark evaluates how well a QPU preserves the algorithmic signal as LR-QAOA depth increases along a linear qubit chain.
This topology isolates the impact of circuit depth from layout constraints.
We use the 99.73\% confidence threshold from a random sampler (Sec.~\ref{A:sampling}) to define benchmark pass/fail behavior.

Figure~\ref{Fig:1Dsmall}(a) shows approximation ratio \(r\) versus depth \(p\) for 100-qubit WMC problems on IBM QPUs.
Among Eagle-generation devices, \texttt{ibm\_brisbane} outperforms peers, while Heron QPUs (\texttt{ibm\_fez}, \texttt{ibm\_marrakesh}, \texttt{ibm\_torino}) show significantly higher \(r\) and slower decay with depth.

Interestingly, the improvement of \texttt{ibm\_torino}-v1, a firmware-enhanced version of \texttt{ibm\_torino}-v0 designed to reduce crosstalk, is captured by this and subsequent benchmarks. However, the EPLG metric at the time of the experiments does not fully reflect this enhancement, with values of 0.8 for v0 and 1.1 for v1. This suggests that error metrics like EPLG that get information of disjoint sections of a QPU may fall short in capturing broader, system-level performance improvements.
Appendix~\ref{A:1D} present additional results of the 1D-Chain test including the time-stability of this benchmark (Fig.~\ref{Fig:1D}(e)).

Figure~\ref{Fig:1Dsmall}(b) compares 5-qubit results across vendors.
\texttt{iqm\_garnet} and \texttt{ibm\_brisbane} perform comparably, implying similar two-qubit error and coherence in their top five qubits. The top 5 qubits of \texttt{rigetti\_ankaa\_2} and \texttt{rigetti\_ankaa\_3} perform similarly and below the other QPUs. This is primarily due to higher two-qubit gate error rates and the need to execute approximately twice as many two-qubit gates compared to other devices.
\texttt{originq\_wukong} maintains coherent output up to \(p=4\), but quickly degrades, consistent with its short decoherence time, \(T_2=5.72 \mu s\).  These results illustrate the limitations of EPLG as a system-wide metric, while for small benchmarks, \texttt{ibm\_marrakesh} outperform the other QPUs in agreement with the lowest reported EPLG (0.4\%), yet does not outperform \texttt{ibm\_fez} at 100 qubits.

Figure~\ref{Fig:1Dsmall}(c) demonstrates the largest depth tested to date: LR-QAOA on \texttt{ibm\_fez} up to \(p=10{,}000\).
Coherent behavior survives until \(p \approx 300\), then relaxes into a noise-dominated regime where \(r\) falls below the random threshold.
This pushes execution beyond 990,000 two-qubit gates, illustrating that some QPUs can already execute deep circuits.

The 1D-chain benchmark establishes a depth-centric baseline for device performance.
It highlights generational improvements and serves as a reference for system quality on the FC benchmark. Extended results of this benchmark are presented in Sec.~\ref{A:1D}.

\subsection{Native layout (NL) benchmark}

\begin{figure*}[!tbh]
\centering
\includegraphics[width=18cm]{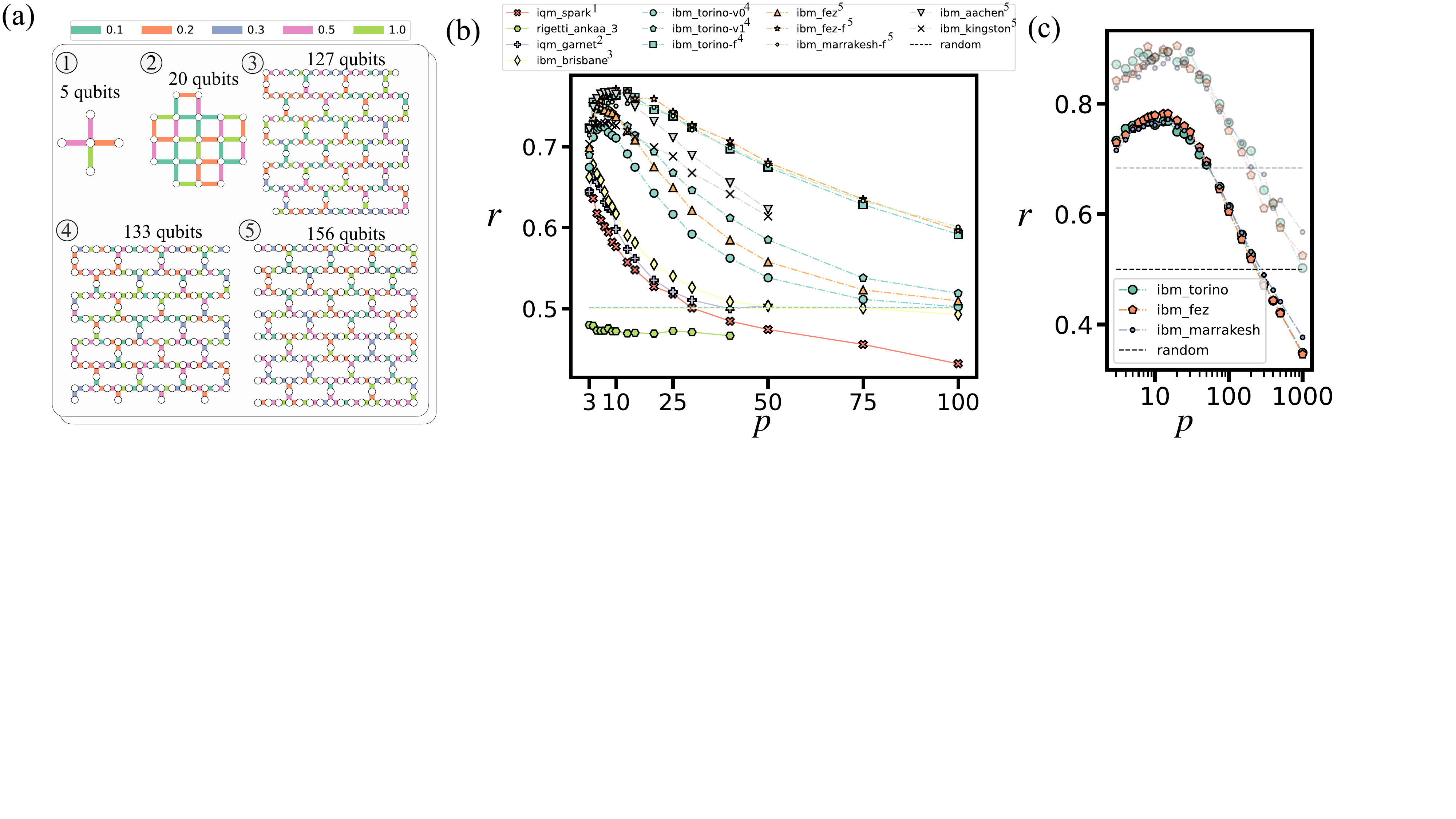}
\caption{\label{Fig:HEsmall} NL-based benchmarking using LR-QAOA for WMC problems on \texttt{iqm\_spark}, \texttt{iqm\_garnet}, \texttt{rigetti\_ankaa\_3}, \texttt{ibm\_brisbane}, \texttt{ibm\_torino}, and \texttt{ibm\_fez}/\texttt{ibm\_marrakesh}/\texttt{ibm\_aachen}/\texttt{ibm\_kingston}. 
Problem sizes (with number of edges in parentheses) are: 5 (4), 20 (30), 82 (138), 127 (144), 133 (150), and 156 (176).
(a) QPU layouts used in each experiment. 
Nodes represent qubits, and edges represent native two-qubit connectivity. 
Edge colors denote the WMC instance weights, randomly chosen from \{0.1, 0.2, 0.3, 0.5, 1\}, as shown at the top.
(b) Approximation ratio versus number of LR-QAOA layers $p$ for each processor.
The dashed horizontal line indicates the random-sampler baseline.
Labels with an 'f' suffix, e.g., \texttt{ibm\_marrakesh}-f, denote experiments using fractional gates.
(c) Scaling experiment with up to \(p = 1{,}000\) LR-QAOA layers using fractional gates.
Semitransparent points mark the sample with the maximum approximation ratio observed.}
\end{figure*}

The NL benchmark stresses system-wide performance by employing all qubits and native couplers in each device’s layout.  
A QPU is considered to pass at a given depth $p$ if the observed approximation ratio $r$ exceeds the 99.73\% confidence threshold of the random-sampler baseline (Sec.~\ref{A:sampling}).  
Among all tested platforms, only \texttt{rigetti\_ankaa\_3} fails to pass the benchmark at any depth. The best-performing devices are \texttt{ibm\_fez}, \texttt{ibm\_marrakesh}, and \texttt{ibm\_torino}, all of which support fractional gates. Interestingly, \texttt{ibm\_aachen} achieves a comparable performance peak despite lacking fractional gate support, implying that it could potentially outperform the other devices once fractional gates become available on this QPU. 

Figure~\ref{Fig:HEsmall}(b) shows the approximation ratio $r$ versus LR-QAOA depth $p$.  
Heron-generation IBM QPUs consistently outperform Eagle-generation devices.  
Within this group, \texttt{ibm\_fez} achieves the best performance, reaching $r = 0.771$ at $p = 10$ using fractional gates.  
Fractional gates provide a clear advantage: they halve the number of two-qubit gates and depth required to implement each LR-QAOA layer, improving effective performance.
\texttt{ibm\_fez}, \texttt{ibm\_marrakesh}, and \texttt{ibm\_torino} all benefit from this support.  
Among devices without fractional support, \texttt{ibm\_aachen} leads with $r = 0.767$ at $p = 9$.  

IQM devices show generation-over-generation improvement. \texttt{iqm\_garnet} maintains coherent signal up to $p = 50$, outperforming \texttt{iqm\_spark}, which drops below the random region by $p = 30$, consistent with thermalization.  
\texttt{rigetti\_ankaa\_3} fails to pass the benchmark at any depth, with performance remaining below the random-sampler threshold throughout.

Figure~\ref{Fig:HEsmall}(c) shows LR-QAOA up to $p = 1{,}000$ on \texttt{ibm\_fez} and \texttt{ibm\_marrakesh}, corresponding to 176,000 two-qubit ZZ gates.  
Coherent signal is retained up to $p = 200$, which involves a two-qubit depth of 600 and time duration of 40.8~$\mu$s.  
Beyond that, performance degrades, consistent with the saturation and decay observed in the 1D-chain benchmark.  

These results validate the NL benchmark as a tool for capturing architectural variation, coherence management, and calibration quality under full-system load.  Extended results of this benchmark are presented in Sec.~\ref{A:HE}. 
Where NL tests expose coherence limits under realistic device topology, the next section isolates circuit depth and routing stress in FC problems.

\subsection{Fully connected (FC) benchmark}

\begin{figure*}[!tbh]
\centering
\includegraphics[width=17cm]{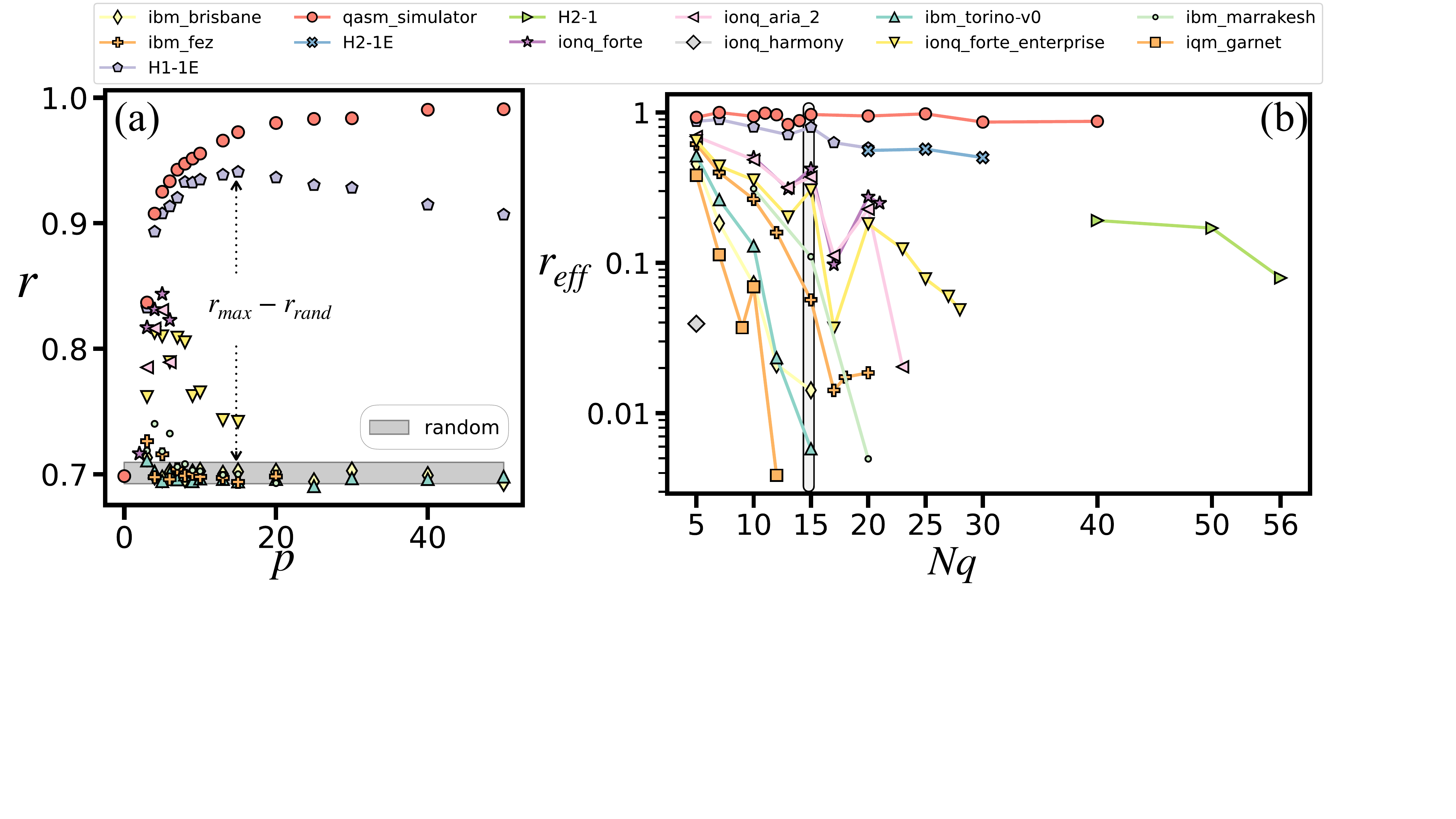}
\caption{\label{Fig:FC}Fully connected (FC) benchmarking using LR-QAOA for WMC problems with 5 to 56 qubits. 
(a) Approximation ratio $r$ versus LR-QAOA depth $p$ for a 15-qubit instance. 
Shaded region denotes the 99.73\% confidence interval of a random sampler. 
(b) Effective approximation ratio $r_{\mathrm{eff}}$ versus problem size for multiple QPUs. 
Each point reflects the best-performing depth not within the random region.}
\end{figure*}

The FC benchmark imposes the highest stress by combining maximal depth and full-qubit connectivity, making it the toughest test of QPU coherence and routing capabilities.  
A device passes if its approximation ratio $r$ exceeds the 99.73\% confidence interval of the random sampler baseline (Sec.~\ref{A:sampling}) at any p. Since the approximation ratio achieved by random guessing increases with problem size, we normalize the approximation ratio to quantify the gain over random guessing using Eq.~\ref{Eq:reff}, and denote this normalized metric as $r_{\text{eff}}$.

The benchmark corresponds to the fully connected topology in Fig.~\ref{Fig:Scheme}(a), which requires $N_{ZZ} = p N_q (N_q - 1)/2$ two-qubit $ZZ$ gates in QPUs with full connectivity between their qubits and 3 times $N_{ZZ}$ for fixed-layout QPUs using a SWAP network encoding strategy (See \ref{A:full_graph}).   
Only a subset of platforms sustain coherent performance at scale.  
Among these, \texttt{quantinuum\_H2-1} executed the largest certified instance: 56 qubits at $p=3$, totaling 4,620 $ZZ$ gates.  
\texttt{ibm\_fez} achieved $r = 0.732 \equiv r_{\text{eff}} = 0.018$ at $N_q = 20$, with 1,710 CZ gates after SWAP encoding insertion.  
\texttt{ionq\_forte\_enterprise} passed at 28 qubits with $r = 0.770 \equiv r_{\text{eff}} = 0.049$.

Figure~\ref{Fig:FC}(a) shows LR-QAOA results on a 15-qubit FC instance, illustrating how noise suppresses signal growth with increasing $p$.  
The \texttt{quantinuum\_H1-1E} emulator closely tracks the noiseless simulator until $p \approx 10$, after which the signal saturates.  
The maximum approximation ratio in hardware is $r=0.94$ at $p=15$, corresponding to $p_{\text{eff}} = 15$. For a performance comparison of the \texttt{quantinuum\_H1-1E} emulator and the real QPU \texttt{quantinuum\_H1-1}, see Sec.~\ref{A:Quantinuum}.

Among IonQ QPUs, \texttt{ionq\_aria\_2}, \texttt{ionq\_forte}, and \texttt{ionq\_forte\_enterprise} reach similar $r_{\max}$ between 0.81 and 0.84 at $p = 5$. Among IBM devices, only Heron QPUs pass the FC benchmark at or above 20 qubits.  
\texttt{ibm\_fez} shows the best results, followed by \texttt{ibm\_marrakesh} at $N_q \leq 15$ with $r_{\max} = 0.74$.  
Eagle-generation QPUs like \texttt{ibm\_brisbane} fail to pass beyond $N_q = 16$.  

Figure~\ref{Fig:FC}(b) summarizes performance across problem sizes using the effective approximation ratio $r_{\text{eff}}$.  
Each point reports the highest $r$ that remains above the statistical noise threshold.  
This visualization confirms that only the Quantinuum family sustains coherent output above 30 qubits.  
\texttt{ibm\_fez} passes up to $N_q = 20$, while other Heron QPUs trail off earlier. \texttt{iqm\_garnet} passes the test up to $N_q = 12$.
\texttt{ionq\_aria\_2}, \texttt{ionq\_forte}, and \texttt{ionq\_forte\_enterprise} demonstrate similar performance. Among them, \texttt{ionq\_forte\_enterprise} has the largest successful run for the 28-qubit problem at \(p=3\) involving 1,134 two-qubit gates.

In Fig.~\ref{Fig:FC}(b), degradation trends reveal scaling limits across platforms. IonQ QPUs deliver similar $r_{eff}$ values up to 20 qubits, but for larger cases, \texttt{ionq\_forte} stops to give results above the random threshold at $N_q=21$ and \texttt{ionq\_aria\_2} at $N_q=23$. 

Quantinuum systems combine a fully connected layout and low gate errors, allowing them to pass the random threshold with their maximum number of qubits, involving up to 4,620 two-qubit gates. This allows \texttt{quantinuum\_H2-1} to reach beyond the exact classical simulation threshold ($N_q \gtrsim 50$)~\cite{DERAEDT201947}.  
The FC experiments yield $r =0.849 \equiv r_{eff}=0.170$ for 50-qubit case and $r =0.872 \equiv r_{eff} = 0.079$ for the 56-qubit case, making them the largest certified QAOA runs to date.

IonQ and Quantinuum QPUs benefit from an all-to-all connectivity that avoids SWAP network encoding overhead. For instance, the 20-qubit problem requires 190 $ZZ$ gates per layer while IBM layouts require 570 CZ gates per layer due to routing (Sec.~\ref{A:full_graph}).  
However, long gate times (600 $\mu$s to 2000 $\mu$s) and lack of parallelism make trapped ion QPUs orders of magnitude slower in wall-clock time.  
See Sec.~\ref{A:time} for a runtime breakdown.

These results show that the FC benchmark separates QPUs not just by noise levels but by architecture and system scaling.  
\texttt{quantinuum\_H2-1}, \texttt{ibm\_fez}, and \texttt{ionq\_forte\_enterprise} represent the current frontier of QAOA execution in terms of both circuit size and algorithmic coherence.  
The random-region threshold provides a clear discriminator for evaluating performance under extreme algorithmic load.

Together, these benchmarks define a performance map that distinguishes hardware by coherence limits, architectural layout, and generational progress. We now synthesize these insights in Table~\ref{Tab:HE}.

\begin{table}[h!]\caption{\label{Tab:HE}Maximum approximation ratio achieved by each QPU across the three benchmark classes. 
For the 1D-chain column, the two values correspond to 5- and 100-qubit instances, respectively. 
Values prefixed with ‘f’ indicate experiments executed using fractional gates. Last column represent the maximum $N_q$ for which a successful experiment is achieved. 
}
\begin{tabular}{ c|c|c|c|c|c|}
 \cline{2-5}
 & \multicolumn{4}{c|}{Graph} \\
 \cline{1-5}
 \multicolumn{1}{ |c| }{QPU} & 1D-Chain & NL & FC$_{N_q=20}$ & FC$_{N_q/r_{eff}}$\\
 \hline
 \multicolumn{1}{ |c| }{\texttt{ibm\_brisbane}} & 0.84/0.756  & 0.678 & 0.000 & 16/0.009\\ 
 \hline
  \multicolumn{1}{ |c| }{\texttt{ibm\_sherbrooke}} & -/0.721  & - & -& -\\ 
 \hline
   \multicolumn{1}{ |c| }{\texttt{ibm\_kyiv}} & -/0.723  & - & -& -\\ 
 \hline
   \multicolumn{1}{ |c| }{\texttt{ibm\_nazca}} & -/0.673  & - & -& -\\ 
 \hline
    \multicolumn{1}{ |c| }{\texttt{ibm\_kyoto}} & -/0.662  & - & -& -\\ 
 \hline
    \multicolumn{1}{ |c| }{\texttt{ibm\_osaka}} & -/0.674  & - & -& -\\
 \hline
    \multicolumn{1}{ |c| }{\texttt{ibm\_brussels}} & -/0.719  & - & -& -\\
 \hline
    \multicolumn{1}{ |c| }{\texttt{ibm\_strasbourg}} & -/0.711  & - & -& -\\
\hline
  \multicolumn{1}{ |c| }{\texttt{ibm\_torino}-v0} & -/0.728 & 0.724 & 0.000 & 15/0.006 \\

\hline

 \multicolumn{1}{ |c| }{\texttt{ibm\_torino}-v1} & -/0.760 & 0.746 & - & - \\
     \multicolumn{1}{ |c|  }{
  } &  & f0.773 &  &  \\
\hline
 \multicolumn{1}{ |c|  }{\texttt{ibm\_fez}} & 0.87/0.776 & 0.751 & 0.020 & 20/0.020 \\
  \multicolumn{1}{ |c|  }{
  } & -/f0.808 & f0.782 &  &  \\
 \hline
  \multicolumn{1}{ |c|  }{\texttt{ibm\_marrakesh}} & 0.92/0.773 & f0.772 & 0.004 & 20/0.004 \\
 \hline
  \multicolumn{1}{ |c|  }{\texttt{ibm\_aachen}} & - & 0.767 & - & -\\
 \hline
  \multicolumn{1}{ |c|  }{\texttt{ibm\_kingston}} & - & 0.731 & - & - \\

\hline
 \multicolumn{1}{ |c|  }{\texttt{H1-1E}} & - & - & 0.582 & 20/0.582 \\

 \hline
 \multicolumn{1}{ |c|  }{\texttt{quantinuum\_H2-1}} & - & - & 0.555 &  56/0.082\\

  \hline
 \multicolumn{1}{ |c|  }{\texttt{ionq\_aria\_2}} & - & - & 0.227 & 23/0.020\\

   \hline
 \multicolumn{1}{ |c|  }{\texttt{ionq\_forte}} & - & - & 0.269 & 21/0.250\\

   \hline
 \multicolumn{1}{ |c|  }{\texttt{ionq\_forte\_enterp}} & - & - & 0.182 & 28/0.049\\

   \hline
 \multicolumn{1}{ |c|  }{\texttt{iqm\_spark}} & - & 0.643 & - & - \\

    \hline
 \multicolumn{1}{ |c|  }{\texttt{iqm\_garnet}} & 0.83/- & 0.664 & - & 12/0.004 \\
     \hline
 \multicolumn{1}{ |c|  }{\texttt{rigetti\_ankaa\_2}} & 0.61/- & - & - & - \\
 \hline
 \multicolumn{1}{ |c|  }{\texttt{rigetti\_ankaa\_3}} & 0.63/- & - & - & - \\
 \hline
 \multicolumn{1}{ |c|  }{\texttt{originq\_wukong}} & 0.84/- & - & - & - \\

 \hline
\end{tabular}
\end{table}

\section{Discussion and Conclusions}\label{Sec:Conclusions}

We presented a benchmarking protocol for QPUs based on the linear ramp quantum approximate optimization algorithm (LR-QAOA), designed to stress circuit width and depth simultaneously.  
Applied to 24 QPUs from six vendors across three graph topologies, a device passes the benchmark if it achieves an approximation ratio statistically distinguishable from a random sampler.  
The protocol yields informative results even with limited sampling and remains stable over time, with less than 1\% variation over two months.  
To our knowledge, this is the most extensive QPU benchmarking study conducted to date.
This positions LR-QAOA as a candidate for standardization in benchmarking suites for noisy intermediate-scale quantum (NISQ) hardware.

These results demonstrate that LR-QAOA functions as a scalable, architecture-neutral benchmark for identifying algorithmic performance limits in QPUs.
It captures depth resilience, routing overheads, and noise accumulation effects in a unified framework.
This contrasts with fidelity-centric or classically verifiable benchmarks, which are often architecture-specific or restricted to fixed-depth circuits.

As the field progresses toward broader deployment and commercial relevance, standardized benchmarks are essential for tracking device capabilities and guiding user selection.
LR-QAOA provides such a reference point, enabling comparative evaluation across hardware types and generations using interpretable, workload-relevant metrics.

These results support several current key findings:  
(i) Heron devices outperform Eagle-generation QPUs in depth scaling and overall algorithmic robustness.  
(ii) Quantinuum's \texttt{H2-1} is the only platform to successfully pass the 56-qubit FC benchmark.  
(iii) The error-per-layered-gate (EPLG) metric does not consistently predict algorithmic performance, especially at scale.  
(iv) LR-QAOA is stable over time and produces reliable benchmarking outcomes even under limited sampling.  
(v) Performance bottlenecks differ by platform: superconducting QPUs are limited by coherence and crosstalk, while ion-trap QPUs are constrained by gate count and execution time.
Vendors whose devices fail the LR-QAOA test at relevant depths may use this protocol to pinpoint bottlenecks, such as coherence limits, qubit connectivity, or inefficient routing layers, and guide improvements in both control software and hardware design.

For the 1D-chain topology, our main test uses 100-qubit problems on IBM devices.  
\texttt{ibm\_fez} achieves the best approximation ratio: \(r = 0.776\) at \(p_{\text{eff}} = 9\) using 1,782 CZ gates, and \(r = 0.808\) at \(p_{\text{eff}} = 15\) using 1,485 ZZ gates with fractional implementation.  
In the NL benchmark, \texttt{ibm\_aachen} attains \(r = 0.767\) at \(p_{\text{eff}} = 9\) with 2,112 CZ gates, while \texttt{ibm\_fez} reaches \(r = 0.782\) at \(p_{\text{eff}} = 15\) with 2,640 ZZ gates.  
For FC problems, \texttt{ibm\_fez} delivers the best IBM result with \(p_{\text{eff}} = 3\) and 1,710 CZ gates on a 20-qubit instance.  
The largest successful LR-QAOA run is a 56-qubit FC problem on \texttt{quantinuum\_H2-1} with \(p = 3\) and 4,620 ZZ gates.

These benchmarks highlight generational progress across QPU platforms.  
IBM devices show marked improvement from the Eagle to Heron generations, driven by both hardware advances and the adoption of fractional gates.  
However, \texttt{ibm\_marrakesh}, released after \texttt{ibm\_fez} with half its reported EPLG, does not exhibit a clear performance advantage.  
Performance degrades with increasing system size, suggesting that non-local crosstalk or thermal effects limit scalability.  
This underscores the limitations of localized error metrics and the importance of system-level calibration.

Following the release of our initial preprint, the 2-qubit gate time of \texttt{quantinuum\_H2-1} was presented \cite{quantinuum2025}, the two-qubit gate constraint ($N_{ZZ} < 650$) on IonQ technology lifted, being able to run a 35-qubit experiment at \(p=5\) on \texttt{ionq\_forte} involving 2,975 two-qubit gates but not passing the test. Additionally, experimental results on \texttt{ibm\_aachen} and \texttt{ibm\_kingston} were contributed by the quantum computing community.  
Such contributions are especially valuable given the limited accessibility of some QPUs and highlight the role of community-shared experiments in benchmarking.

Among IonQ devices, we observe substantial improvement from \texttt{ionq\_harmony} to \texttt{ionq\_aria\_2}, with similar performance between \texttt{ionq\_aria\_2} and \texttt{ionq\_forte} and a gain in the number of qubits passing the test for \texttt{ionq\_forte\_enterprise}.  
Quantinuum’s \texttt{H1-1} and \texttt{H2-1} perform similarly and consistently outperform other QPUs.  
IQM’s \texttt{iqm\_garnet} improves notably over \texttt{iqm\_spark}.  
For Rigetti, the best five qubits on \texttt{ankaa\_2} and \texttt{ankaa\_3} show similar performance, with no significant gains in the newer generation. \texttt{ankaa\_3} fails to pass the NL benchmark at any tested depth.

Circuit depth is a defining feature of LR-QAOA, highlighting the challenges of executing deep quantum circuits.  
Superconducting platforms (IBM, IQM, Rigetti, OriginQ) offer fast gate times but higher two-qubit error rates.  
Trapped-ion platforms (IonQ, Quantinuum) have lower error rates and full connectivity but are limited by slower gate speeds and lack of parallelism.  
For example, a 25-qubit FC problem with \(p = 100\) and 1,000 shots takes approximately 18,000 seconds on \texttt{ionq\_aria\_2}, compared to 0.5 seconds on \texttt{ibm\_fez}.  
Gate volume limits also vary widely: IBM supports nearly one million two-qubit gates per circuit, IonQ limited this to under 650 at the time of the release of the first version of this manuscript, and Quantinuum ties usage cost mainly to two-qubit gate count.  
The LR-QAOA benchmark reveals these throughput bottlenecks and emphasizes the need to improve both coherence and execution rate for scalable quantum applications.

Several strategies can improve LR-QAOA performance.  
Dynamical decoupling~\cite{Ezzell_2023} can mitigate decoherence.  
For FC problems on fixed-layout devices, improved circuit transpilation~\cite{ji2024, klaver2024,montanezbarrera2025optimizingqaoacircuittranspilation} and routing strategies~\cite{Nation_2023, montanezbarrera2024c} can reduce overhead and errors.  
Postprocessing measurement has not shown reliable error suppression; if applied, it must be used consistently on the random baseline.  
Changing the initial state, for example via warm-start initialization~\cite{Egger_2021}, introduces bias and invalidates statistical comparisons. In general, these benchmarks are designed to capture the raw performance of a QPU. While many enhancements can be applied in real applications to improve the approximation ratio, they are not desirable in the benchmark context, which focuses solely on whether a QPU remains coherent at a given depth. For a fair comparison, the same set of parameters must be used across all devices.

As QPUs continue to evolve, LR-QAOA provides a scalable, statistically rigorous benchmark to evaluate circuit width, depth, coherence, and runtime bottlenecks.  
Looking forward, benchmarking protocols must evolve to assess hybrid quantum-classical workflows, application-specific workloads, and performance under noise-aware algorithmic strategies.
In particular, tailoring cost Hamiltonians to reflect real-world workloads could provide more targeted application readiness assessments.  
When selecting hardware for depth-intensive quantum algorithms, these insights can inform QPU design trade-offs, calibration priorities, and user expectations.

\section*{Data Availability}
All problem instances and results analyzed in this study are available at: \url{https://github.com/alejomonbar/LR-QAOA-QPU-Benchmarking}.

\begin{acknowledgments}
\vspace{-10pt}

J. A. Montanez-Barrera acknowledges support from the German Federal Ministry of Education and Research (BMBF), the funding program Quantum Technologies - from Basic Research to Market, project QSolid (Grant No. 13N16149).
D. E. Bernal Neira acknowledges the support of the startup grant of the Davidson School of Chemical Engineering at Purdue University.
We acknowledge the support of the Quantum Collaborative for their support and access to IBM Quantum Resources.
The views expressed are those of the authors and do not reflect the official policy or position of IBM or the IBM Quantum team.

This research used resources of the Oak Ridge Leadership Computing Facility for the experiments on Quantinuum QPUs, which is a DOE Office of Science User Facility supported under Contract DE-AC05-00OR22725.

\end{acknowledgments}

\bibliography{References}

\clearpage  
\onecolumngrid  
\appendix

\section*{Supplementary Material}\label{Sec:appendix}

\subsection{Experimental \texorpdfstring{$\Delta_{\gamma,\beta}$}{Delta(gamma,beta)} used}
\label{A:Deltas}

For 1D-chain and native layout (NL) problems, we use a fixed ramp value of $\Delta_{\gamma,\beta} = 1$.  
For fully connected (FC) problems, the value of $\Delta_{\gamma,\beta}$ depends on the problem size.  
For instances with $N_q \le 15$, we use $\Delta_{\gamma,\beta} = 0.63$ across all QPUs.  
For $N_q > 15$, the values used in each experiment are summarized in Table~\ref{Tab:deltas}.

These values were not tuned for performance on individual QPUs.  
Instead, they follow the observation from~\cite[Sec.~III-B]{Montanez-Barrera2024b} that the optimal value of $\Delta_{\gamma,\beta}$ decreases as $N_q$ increases, in order to maintain algorithmic signal.  
In principle, one could define a functional form $\Delta_{\gamma,\beta}(N_q)$ for improved scaling, or perform a parameter sweep to find the optimal value per instance.  
Examples of performance sensitivity to $\Delta_{\gamma,\beta}$ are shown in Fig.~\ref{Fig:1D}(a) and Fig.~\ref{Fig:HE}(b).

\begin{table}[h!]\caption{Backends $\Delta_{\gamma,\beta}$ used.}\label{Tab:deltas}
\begin{tabular}{c|c|c|c|c|c|c|c|c|c|c|c|c|c|c|c|c|c|}
 \cline{2-17}
 & \multicolumn{16}{c|}{Backend} \\
 \cline{2-17}
 & \texttt{ibm\_fez}(\texttt{marrakesh}) & \multicolumn{2}{|c|}{\texttt{ibm\_torino}} & \multicolumn{3}{|c|}{\texttt{ibm\_brisbane}} & \texttt{H1-1E} & \multicolumn{2}{|c|}{\texttt{H2-1E}} & \multicolumn{3}{|c|}{\texttt{H2-1}} &\multicolumn{2}{|c|}{\texttt{ionq\_aria\_2}} &\multicolumn{2}{|c|}{\texttt{qasm\_simulator}}\\
 \hline
 \multicolumn{1}{ |c| }{$N_q$} & $> 15$ & 17 & 20 & 16 & 17 & 20 & 20 & 25 & 30 & 40 & 50 & 56 & 17 & 20 & 20 & 25\\ \hline

 \multicolumn{1}{ |c|  }{$\Delta_{\beta,\gamma}$} & 0.63 & 0.4 & 0.3 & 0.5 & 0.4 &  0.3 & 0.3 & 0.5 & 0.4& 0.2 & 0.2 & 0.2 & 0.63 & 0.3 & 0.3 & 0.4\\

 \hline
\end{tabular}
\end{table}

\subsection{Circuit Gate Counts and Depth Formulas}
\label{A:circuit_scaling}

This section summarizes analytical expressions for two-qubit gate counts and circuit depth as a function of QAOA layer count \( p \) and device architecture.

For 1D-chain problems on IBM and IQM devices using native CZ gates, the total number of two-qubit gates is  
\[
N_{2q} = 2p(N_q - 1), 
\]
and the circuit depth is  
\[
d = 4p.
\]  
Rigetti QPUs use iSWAP-based implementations, resulting in  
\[
N_{2q} = 4p(N_q - 1), \quad d = 8p.
\]

For native layout (NL) problems, the gate count scales with the number of edges \( N_{\text{edges}} \) in the device's native connectivity graph:  
\[
N_{2q} = 2p N_{\text{edges}}.
\]  
The circuit depth depends on the layout type:  
\[
d = 6p \quad \text{(heavy-hex)}, \qquad d = 8p \quad \text{(square lattice)}.
\]

Specific NL gate counts for CZ-based QPUs are:  
\[
N_{2q} = 288p \quad \text{(Eagle)}, \quad 300p \quad \text{(Heron r1)}, \quad 352p \quad \text{(Heron r2)}, \quad 60p \quad \text{(\texttt{iqm\_garnet})}.
\]  
Corresponding depths are \( d = 6p \) for IBM and \( d = 8p \) for IQM devices.

For fully connected (FC) problems, the total number of logical two-qubit gates is  
\[
N_{ZZ} = \frac{p N_q(N_q - 1)}{2}.
\]  
On fixed-layout QPUs, SWAP-based routing introduces a 3x overhead:  
\[
N_{2q} = \frac{3p N_q(N_q - 1)}{2}, \quad d = 3pN_q.
\]  
For all-to-all architectures like \texttt{quantinuum\_H2-1}, gate counts follow the logical requirement, and circuit depth is  
\[
N_{2q} = \frac{p N_q(N_q - 1)}{2}, \quad d = \frac{p N_q(N_q - 1)}{8}.
\]
\subsection{Random Sampling Limit}
\label{A:sampling}

To determine whether a QPU result is statistically meaningful, we compare the approximation ratio obtained from LR-QAOA with that from a random sampler.
Noise typically causes degradation in QPU output, and we classify the resulting behavior into three regimes.
In the first regime, the approximation ratio is above the 99.73\% confidence interval of the random sampler, indicating a meaningful algorithmic signal.
In the second regime, the approximation ratio falls within this interval, suggesting that the output is statistically indistinguishable from a fully mixed distribution.
In the third regime, the QPU output yields approximation ratios consistently below the random threshold, often due to decoherence or state relaxation.

This latter regime occurs in high-depth circuits, such as the 5-qubit NL experiment on \texttt{iqm\_spark} at \(p = 100\) (800 two-qubit gates), or the 100-qubit FC experiment on \texttt{ibm\_torino} at \(p = 20\), requiring 297,000 two-qubit gates.
Figure~\ref{Fig:torino100FC} illustrates this degradation in \texttt{ibm\_torino}, where results fall below the random region.

\begin{figure}[!tbh]
\centering
\includegraphics[width=18cm]{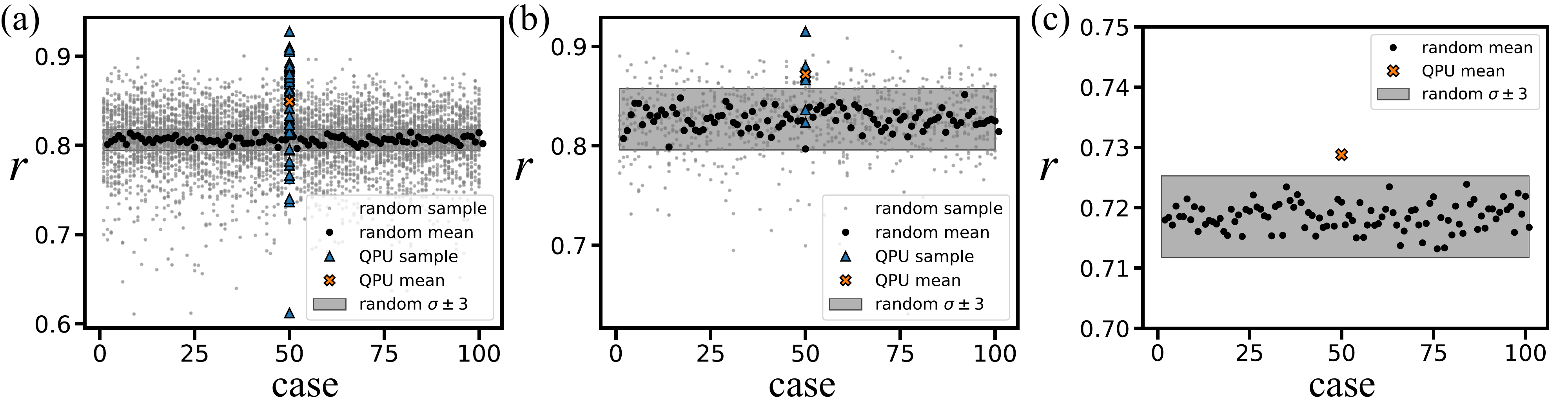}
\caption{\label{Fig:sampling} Comparison of LR-QAOA approximation ratios to random sampler baselines for fully connected WMC problems.
(a) \texttt{quantinuum\_H2-1}, 50-qubit problem, 50 samples, \(p = 4\);
(b) \texttt{quantinuum\_H2-1}, 56-qubit problem, 7 samples, \(p = 3\);
(c) \texttt{ibm\_fez}, 20-qubit problem, 1000 samples, \(p = 3\).
Dots represent individual random sample values; dark circles denote their mean; triangles indicate QPU samples; the X marker is the QPU sample mean; and shaded regions show the \(\pm 3\sigma\) random confidence interval.
Samples are not shown in (c) for visibility.}
\end{figure}

\begin{figure*}[h!]
\centering
\includegraphics[width=8cm]{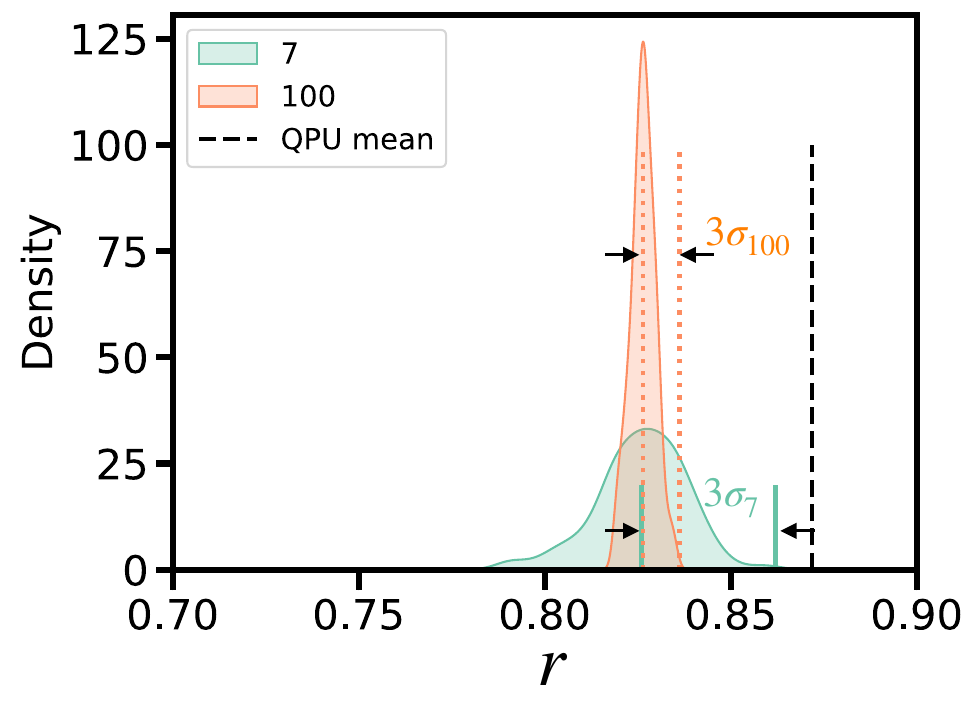}
\caption{\label{Fig:sampling_56} Distribution of approximation ratios for 56-qubit FC WMC problems using 7 and 100 random samples.
The standard deviations \(\sigma_7\) and \(\sigma_{100}\) are computed from 200 repeated samplings.
Dashed line indicates the mean result from \texttt{quantinuum\_H2-1} using 7 samples.}
\end{figure*}

\begin{figure}[!tbh]
\centering
\includegraphics[width=9cm]{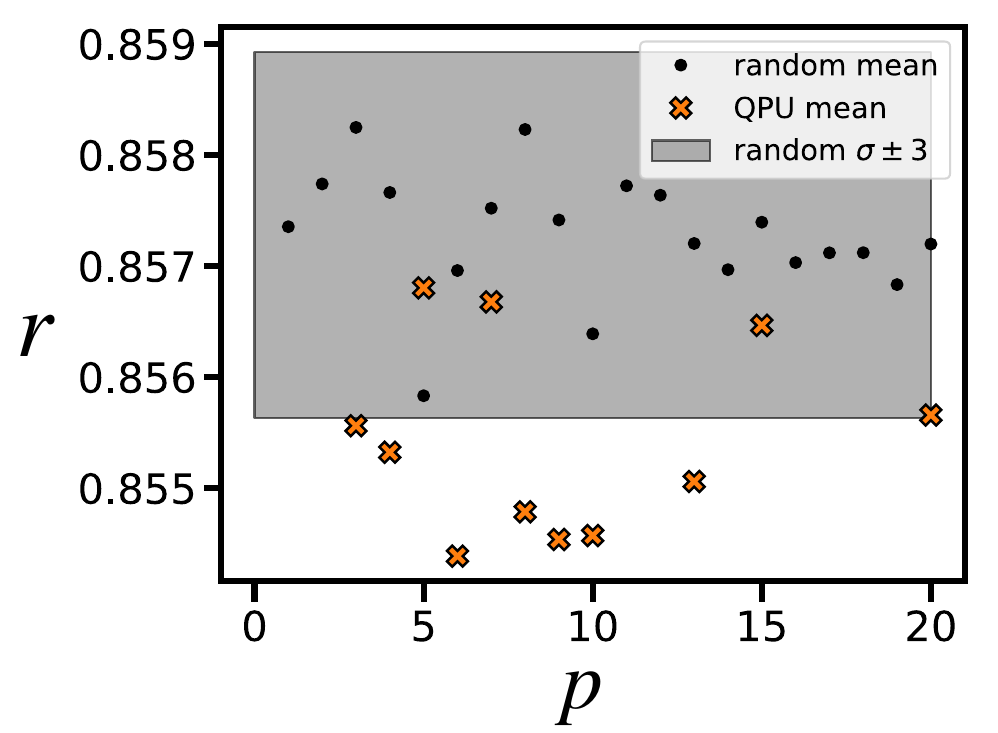}
\caption{\label{Fig:torino100FC} Approximation ratios from LR-QAOA on \texttt{ibm\_torino} for a 100-qubit fully connected WMC problem, with \(p\) ranging from 3 to 20.
Each case represents a different random instance.
Visualization follows the same convention as Fig.~\ref{Fig:sampling}.}
\end{figure}

Figure~\ref{Fig:sampling} presents FC benchmark results comparing QPU samples to random sampling baselines.
In Fig.~\ref{Fig:sampling}(a), \texttt{quantinuum\_H2-1} is used to run a 50-qubit problem with 50 samples at \(p = 4\); in Fig.~\ref{Fig:sampling}(b), only 7 samples are used for a 56-qubit problem at \(p = 3\).
As expected, fewer samples broaden the confidence interval, but the LR-QAOA result still remains distinguishable from random sampling.
This is further shown in Fig.~\ref{Fig:sampling_56}, where the standard deviation increases by a factor of 3.68 when reducing from 100 to 7 samples.
Hence, deeper circuits or smaller sample sizes demand more pronounced algorithmic signal to achieve statistical certification.

In contrast, superconducting QPUs such as \texttt{ibm\_fez} allow for lower-cost sampling.
In Fig.~\ref{Fig:sampling}(c), 1000 samples are used for a 20-qubit problem, producing a narrow random baseline confidence interval.

To determine whether a QPU is performing meaningful optimization, we use the effective approximation ratio:
\begin{equation}\label{Eq:reff}
    r_{\mathrm{eff}} = \frac{r_{\mathrm{max}} - r_{\mathrm{rand}}}{1 - r_{\mathrm{rand}}},
\end{equation}
where \(r_{\mathrm{max}}\) is the maximum observed approximation ratio from the QPU, and \(r_{\mathrm{rand}}\) is the mean approximation ratio of a random sampler plus three standard deviations over 100 sampled subsets.
If \(r_{\mathrm{eff}} > 0\), the QPU result exceeds the random threshold with 99.73\% confidence and is considered statistically meaningful.

\subsection{1D-chain Benchmarking}
\label{A:1D}

The 1D-chain benchmark provides a practical diagnostic of coherence and gate fidelity along a device's longest physical qubit chain.  
This is motivated by the fact that fully connected (FC) interactions can be constructed from a linear qubit chain using the SWAP strategy described in Sec.~\ref{A:full_graph}, with depth scaling as $O(N_q)$.

\begin{figure*}[!tbh]
\centering
\includegraphics[width=18cm]{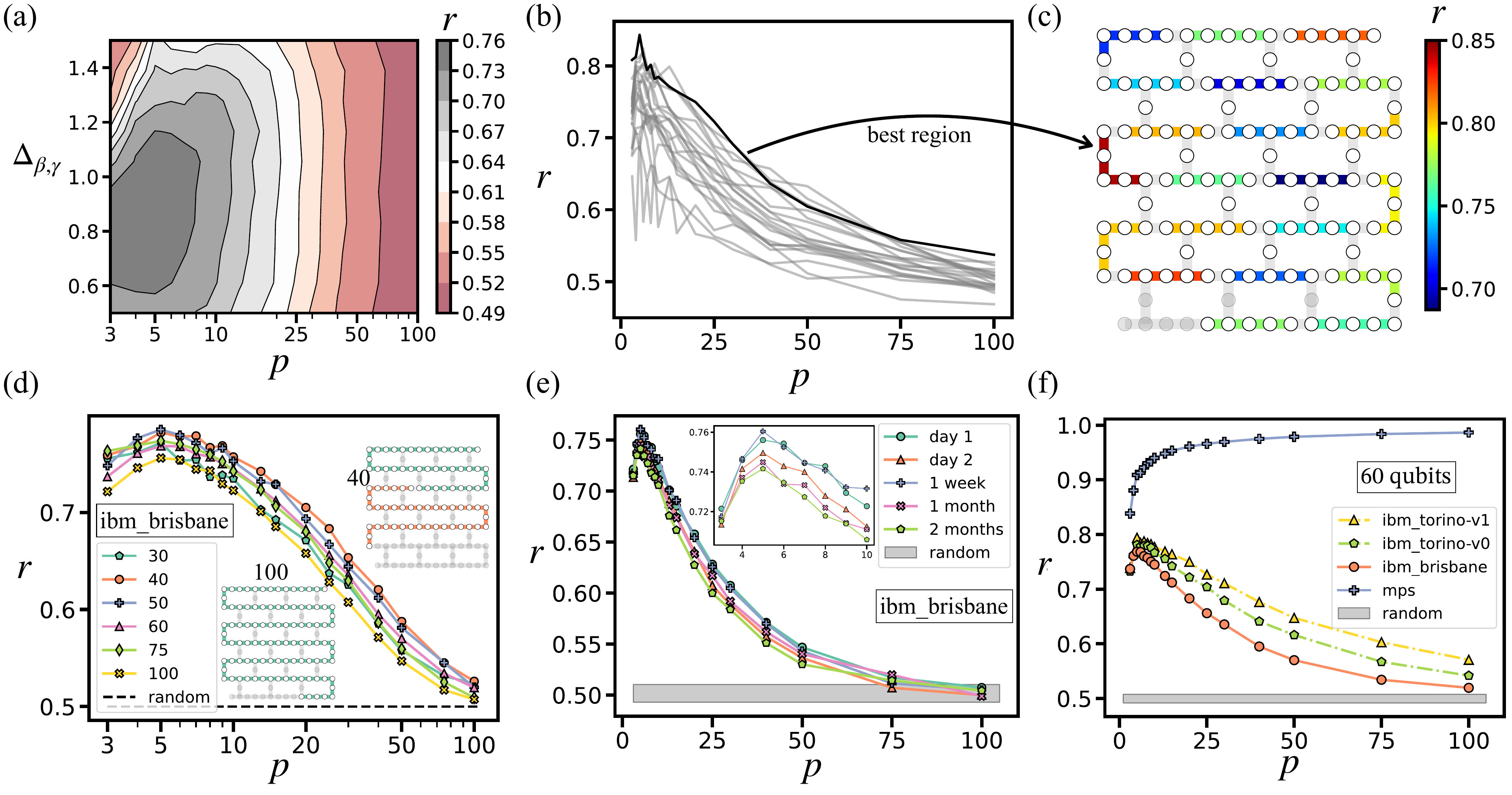}
\caption{\label{Fig:1D}1D-chain WMC problems using LR-QAOA for $p=3$ to 100 and problem sizes between 19 and 100 qubits on IBM, IQM, and Rigetti devices.  
(a) Performance diagram on \texttt{ibm\_brisbane} (100 qubits).  
(b) Approximation ratio for a 5-qubit problem on different device sections.  
(c) Layout map showing results from (b), with edge colors indicating approximation ratio.  
(d) Approximation ratio versus $p$ for different $N_q$; insets show selected chains.  
(e) Time stability of a 100-qubit run on different days.  
(f) Comparison between experimental results and an MPS simulation for $N_q = 60$.}
\end{figure*}

Figure~\ref{Fig:1D} presents LR-QAOA results on IBM QPUs, \texttt{iqm\_garnet}, and \texttt{rigetti\_ankaa}.  
Figure~\ref{Fig:1D}(a) shows a performance diagram of the approximation ratio as a function of $\Delta_{\gamma,\beta}$ and $p$ for a 100-qubit WMC instance on \texttt{ibm\_brisbane}.  
This analysis motivates the choice of $\Delta_{\gamma,\beta} = 1$ for all subsequent 1D-chain experiments.

Figure~\ref{Fig:1D}(b) shows the approximation ratio for a 5-qubit problem evaluated across different sections of \texttt{ibm\_brisbane}.  
The section with the highest approximation ratio is highlighted in black.  
Figure~\ref{Fig:1D}(c) maps these results to physical qubit locations, illustrating section-wise performance across the 109-qubit main diagonal of the device.  
This section selection procedure is used consistently across 1D-chain and FC benchmarks.

Figure~\ref{Fig:1D}(d) shows approximation ratio versus depth $p$ for problem sizes ranging from 30 to 100 qubits.  
Performance generally improves with system size, though a small gap appears between the 30- and 100-qubit cases.  
This is consistent with LR-QAOA behavior, which reflects the average fidelity of the qubits involved.

Figure~\ref{Fig:1D}(e) reports on a 100-qubit instance executed on \texttt{ibm\_brisbane} over multiple days.  
The standard deviation at peak approximation ratio is $\sigma = \pm 0.007$ with a mean value $\mu = 0.750$, yielding a coefficient of variation of $\sigma / \mu = 1\%$.  
This confirms the temporal stability and reproducibility of the benchmark.

Figure~\ref{Fig:1D}(f) compares experimental results for a 60-qubit WMC instance on \texttt{ibm\_brisbane}, \texttt{ibm\_torino}-v0, and \texttt{ibm\_torino}-v1 with a Matrix Product State (MPS) simulator.  
Results show clear generational improvement, with firmware upgrades (v1) yielding measurable gains.  
The MPS simulation provides a reference for coherent evolution in low-dimensional circuits and confirms expected LR-QAOA trends under idealized conditions.

\subsection{Correlation in LR-QAOA}
\label{A:Correlation}

\begin{figure*}[h!]
\centering
\includegraphics[width=18cm]{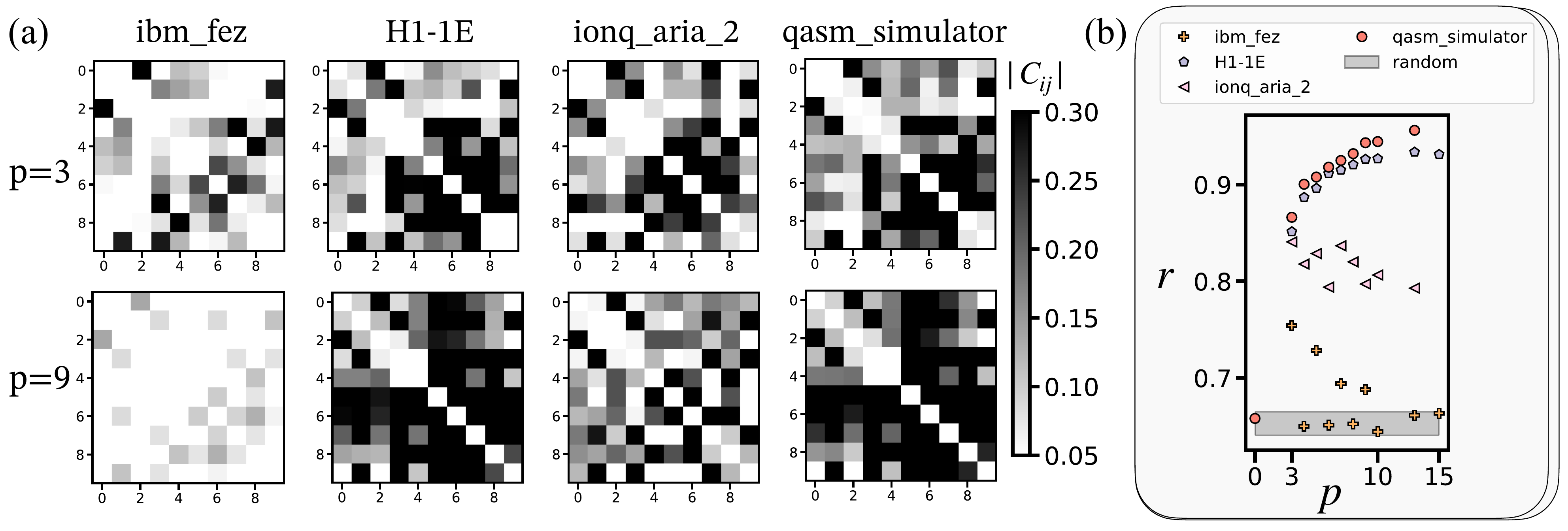}
\caption{\label{Fig:Correlation} 
(a) Correlation matrices \( |C_{ij}| \) from 10-qubit FC WMC problems at \(p = 3\) (top row) and \(p = 9\) (bottom row) on different QPUs.  
(b) Approximation ratio \(r\) versus number of LR-QAOA layers \(p\) for the same problem instances.}
\end{figure*}

Figure~\ref{Fig:Correlation} shows the pairwise correlation structure of bitstring samples for a 10-qubit FC WMC problem solved via LR-QAOA on different QPUs at depths \(p = 3\) and \(p = 9\).  
Correlations are computed as
\begin{equation}
    |C_{ij}| = \left| \frac{1}{N} \sum_{n=1}^N s^n_i s^n_j \right|,
\end{equation}
where \(s^n_i \in \{0,1\}\) is the value of qubit \(i\) in sample \(n\), and \(N\) is the total number of samples.  
This metric captures statistical dependencies in the output distribution that reflect the entangled structure generated by the LR-QAOA circuit.

Figure~\ref{Fig:Correlation}(a) shows the correlation matrices produced by a noiseless simulator (\texttt{qasm\_simulator}), \texttt{quantinuum\_H1-1E}, \texttt{ionq\_aria\_2}, and \texttt{ibm\_fez}.
At \(p = 3\), all platforms retain some nontrivial correlations, although the ideal structure is most closely preserved by \texttt{quantinuum\_H1-1E}.  
By \(p = 9\), both \texttt{ionq\_aria\_2} and \texttt{ibm\_fez} exhibit significant loss of correlation, while the emulator continues to approximate the ideal evolution.  
This behavior reflects coherence degradation and sampling noise, with IBM showing earlier onset of decoherence compared to the trapped-ion platforms.

Figure~\ref{Fig:Correlation}(b) reports the corresponding approximation ratios as a function of depth \(p\), corroborating the correlation-based interpretation.

\subsection{Native Layout Benchmarking}
\label{A:HE}

\begin{figure*}[!tbh]
\centering
\includegraphics[width=18cm]{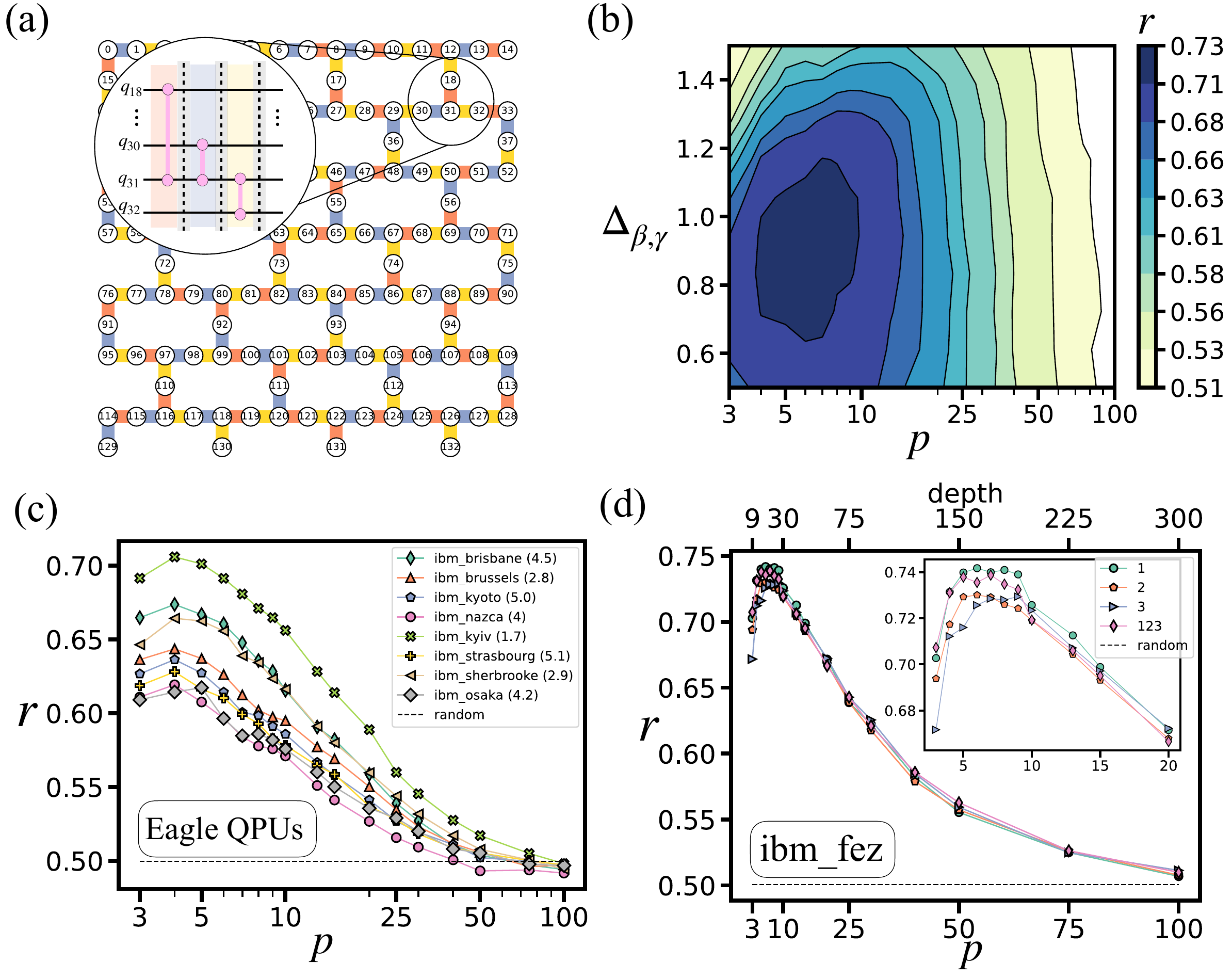}
\caption{\label{Fig:HE} 
Native layout benchmarking using LR-QAOA for WMC problems on \texttt{iqm\_garnet}, \texttt{ibm\_brisbane}, \texttt{ibm\_torino}, and \texttt{ibm\_fez}, with $N_q$ ranging from 20 to 156.  
(a) Layout of \texttt{ibm\_torino}, with color-coded two-qubit gate groups and a 2-qubit depth inset.  
(b) Performance diagram for the NL benchmark on \texttt{ibm\_torino}.  
(c) Approximation ratio vs. $p$ across Eagle devices (EPLG in parentheses).  
(d) Robustness to random WMC weight seeds on \texttt{ibm\_fez}.  
The dashed line indicates the statistical threshold for random sampling.}
\end{figure*}

Figure~\ref{Fig:HE} presents results for LR-QAOA applied to native layout (NL) WMC problems, where all qubits and native couplers of each device are used.  
Figure~\ref{Fig:HE}(a) shows the qubit layout of a 133-qubit Heron r1 device (\texttt{ibm\_torino}).  
To implement LR-QAOA, the circuit is decomposed into three sublayers of two-qubit gates, each executing a similar number of operations in parallel.  
Figure~\ref{Fig:HE}(b) displays the performance diagram for \texttt{ibm\_torino}, mapping the approximation ratio as a function of depth $p$ and parameter scale $\Delta_{\beta,\gamma}$.  
In the ideal case, $r$ approaches 1, as seen in simulated benchmarks (Fig.~\ref{Fig:PD}(a)).  
We use these results to select $\Delta_{\beta,\gamma} = 1$ for all NL experiments.

Figure~\ref{Fig:HE}(c) compares IBM Eagle devices on NL problems.  
Among them, \texttt{ibm\_kyiv}, which reports the lowest EPLG at the time of testing, achieves the highest approximation ratio.  
However, the observed performance across other devices does not consistently align with EPLG values.  
All tested Eagle devices show an effective depth $p_{\text{eff}} = 4$, marking the onset of decoherence-dominated behavior.

Figure~\ref{Fig:HE}(d) reports approximation ratios on \texttt{ibm\_fez} for different WMC instances with randomly seeded edge weights.  
Despite varying weight realizations, performance remains consistent, with a standard deviation of $\pm 0.005$ at peak $r$.  
This supports the benchmark’s robustness: random weight assignments drawn from a fixed value set (0.1, 0.2, 0.3, 0.5, 1.0) yield statistically stable results, removing the need for hand-crafted problem instances.

\begin{figure*}[!tbh]
\centering
\includegraphics[width=16cm]{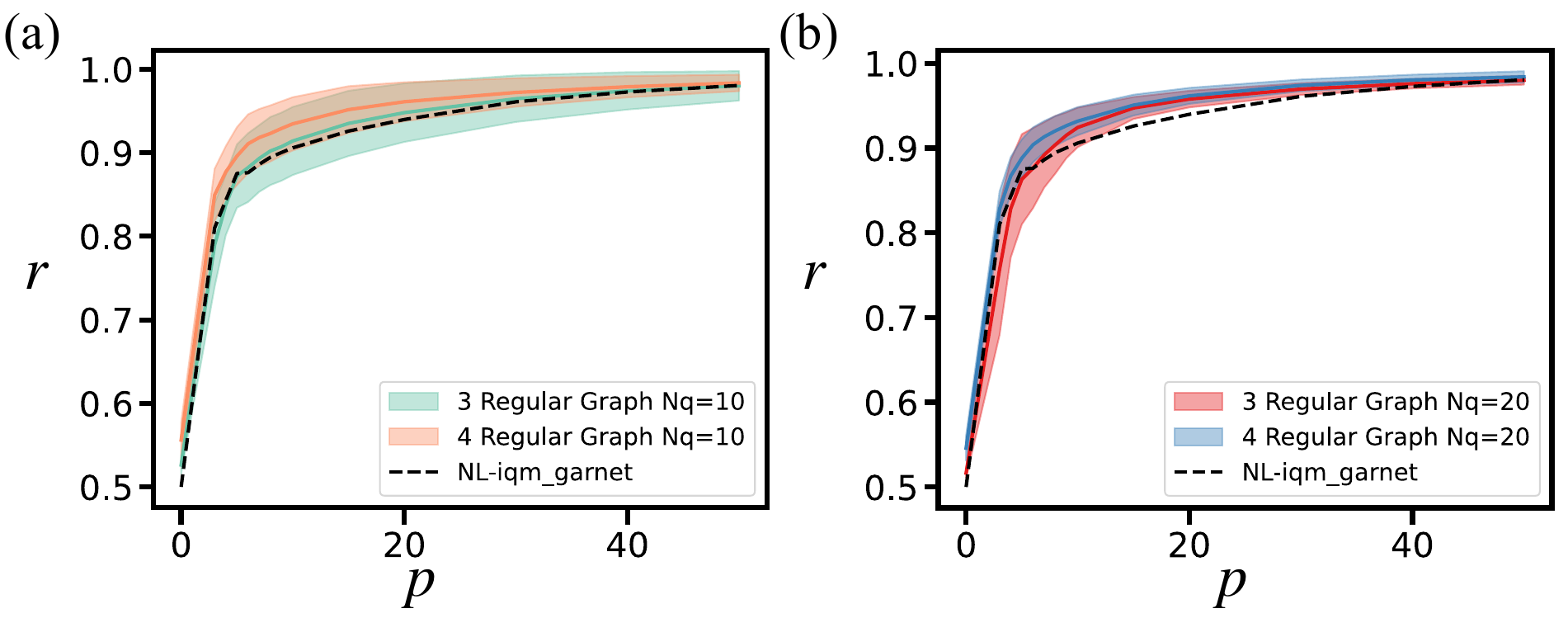}
\caption{\label{Fig:NL_graphs_comp} 
Comparison of LR-QAOA results for 10 random 3- and 4-regular WMC instances on (a) 10-qubit and (b) 20-qubit graphs.  
Shaded bands denote mean $\pm$ standard deviation.  
Dashed lines show ideal convergence curves.}
\end{figure*}

Figure~\ref{Fig:NL_graphs_comp} shows LR-QAOA performance for 3- and 4-regular graphs with 10 and 20 qubits.  
Despite differences in graph structure and problem size, the approximation ratio curves exhibit consistent scaling.  
This reinforces the protocol’s reliability and justifies aggregating QPU results in performance summaries such as Fig.~\ref{Fig:HEsmall}.  
The filled bands represent the standard deviation across 10 randomly sampled instances, while the dashed line reflects ideal LR-QAOA convergence on the same topology.

\subsection{SWAP strategy}
\label{A:full_graph}

\begin{figure*}[!ht]
\centering
\includegraphics[width=16cm]{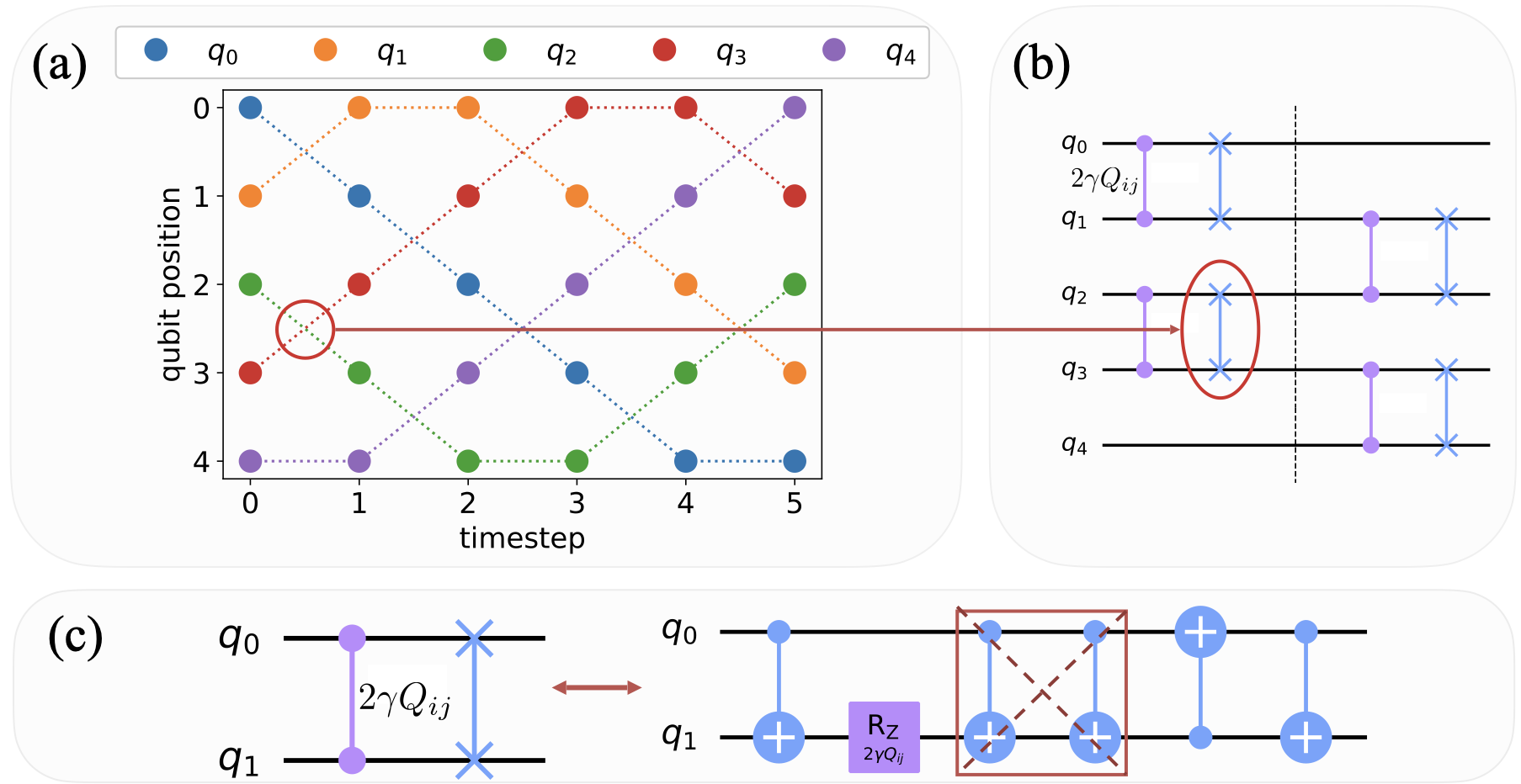}
\caption{\label{swap_strategy} (a) Linear SWAP network strategy. Colored circles represent logical qubits, and the vertical axis corresponds to physical qubit positions.  
(b) Circuit representation of the SWAP network interleaved with two-qubit terms in the QUBO.  
(c) Gate decomposition of SWAP and $RZZ$ gates, showing CNOT cancellation.}
\end{figure*}

For fully connected (FC) problems on fixed-layout QPUs, we map the logical connectivity onto a linear topology using a SWAP network~\cite{Hirata2009, Jin2021}.  
This strategy, illustrated in Fig.~\ref{swap_strategy}(a), implements a complete interaction pattern by permuting logical qubit positions over time.  
The vertical axis denotes physical qubits, and the horizontal axis represents timesteps.  
Colored circles indicate logical qubits, which are routed through SWAP operations as shown by the dotted lines.  
Each SWAP operation corresponds to a gate sequence as depicted in the red oval in Fig.~\ref{swap_strategy}(b).

The linear SWAP network requires $O(N_q(N_q - 1)/2)$ SWAP gates per QAOA layer and increases the circuit depth by a factor proportional to $N_q$.  
For devices that use CNOT as the native two-qubit gate, a SWAP is decomposed into three CNOTs.  
The $RZZ(2\gamma Q_{ij})$ gate, used in the problem unitary, can be decomposed into two CNOTs and a single-qubit $RZ$ rotation, as shown in Fig.~\ref{swap_strategy}(c).  
When these gates are composed together, two CNOTs cancel, reducing the total number of CNOTs required.

Thus, implementing an FC QAOA layer on a fixed-layout QPU requires $N_{2q} = \frac{3p N_q (N_q - 1)}{2}$ CNOT gates.  
This routing overhead makes the FC benchmark particularly demanding for superconducting architectures with limited connectivity.

\subsection{Quantinuum Experiments}
\label{A:Quantinuum}

Access to Quantinuum hardware is governed by a quantum cost metric known as hardware quantum cost (HQC), computed as

\begin{equation}
\mathrm{HQC} = 5 + \frac{(N_{1q} + 10 N_{2q} + 5 N_q)N_{\text{shots}}}{5000},
\end{equation}

where \(N_{1q}\) is the number of single-qubit gates, \(N_{2q}\) is the number of two-qubit gates, \(N_q\) is the number of qubits, and \(N_{\text{shots}}\) is the number of measurement samples.

For LR-QAOA circuits solving fully connected WMC problems, the gate counts are \(N_{1q} = pN_q\) and \(N_{2q} = \frac{pN_q(N_q - 1)}{2}\).  
For example, a 20-qubit instance with \(p = 20\) and 50 shots requires \(\mathrm{HQC} = 390.2\), while a 50-qubit problem with \(p = 3\) and 50 shots requires \(\mathrm{HQC} = 472\).  
Given a budget cap of \(\mathrm{HQC} = 2000\), we allocated this budget toward a large-scale experiment on real hardware (\texttt{quantinuum\_H1-1}) and ran smaller instances on its emulator (\texttt{H1-1E}).

\begin{figure*}[!ht]
\centering
\includegraphics[width=7cm]{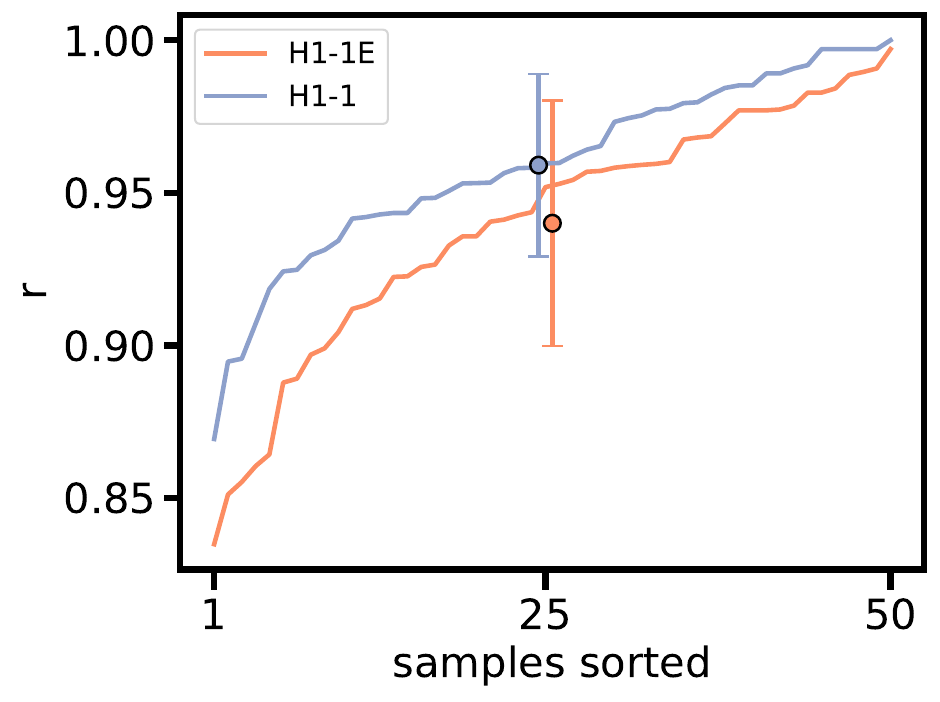}
\caption{\label{Fig:h1_1_vs_h1_1E} Fully connected 20-qubit WMC problem solved using LR-QAOA with \(p = 20\) on \texttt{quantinuum\_H1-1} and its emulator \texttt{H1-1E}. 
Each marker shows a sample’s approximation ratio, sorted by value. 
Error bars denote sample standard deviation.}
\end{figure*}

Figure~\ref{Fig:h1_1_vs_h1_1E} compares results from the real device and emulator on a 20-qubit FC WMC problem at \(p = 20\).  
The plot displays the approximation ratios of 50 samples sorted in ascending order.  
\texttt{H1-1E} yields lower approximation ratios on average, suggesting that its noise model is not overly optimistic relative to the actual \texttt{H1-1} hardware.

\subsection{Execution Time Projections}
\label{A:time}

\begin{figure}[h!]
\centering
\includegraphics[width=8cm]{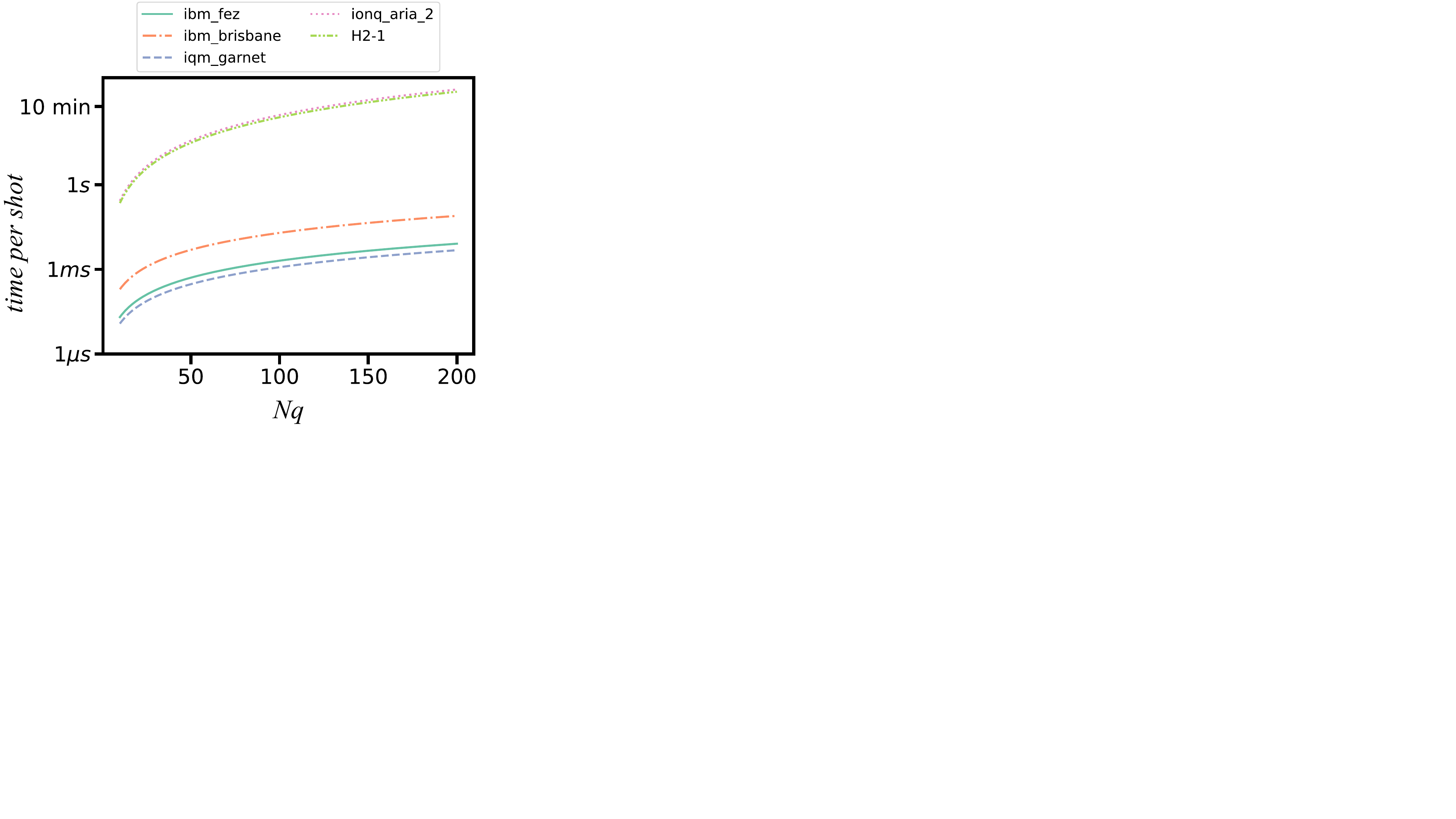}
\caption{\label{Fig:time-QAOA} 
Projected execution time for fully connected LR-QAOA problems with \(p = N_q\) on three representative QPUs.  
Times reflect the cumulative two-qubit gate duration only.}
\end{figure}

Execution time is an important constraint in quantum algorithms, especially for deep circuits with large qubit counts.  
Figure~\ref{Fig:time-QAOA} compares projected runtime estimates for a fully connected LR-QAOA circuit using \(p = N_q\) layers on three representative QPUs.  
Estimates are based solely on the cumulative two-qubit gate durations, ignoring classical overhead and measurement latency.

For superconducting devices such as \texttt{ibm\_fez}, execution is rapid even at large scales, requiring only 0.5~ms to run a 200-qubit problem with 200 layers.  
In contrast, trapped-ion QPUs (\texttt{ionq\_aria\_2} and \texttt{quantinuum\_H2-1}) require tens of minutes for the same configuration due to slower gate speeds and limited parallelism.

These results highlight that while ion-trap platforms maintain high fidelity and full connectivity, their runtime costs can limit practical circuit depth, especially when high shot counts are needed for statistical confidence.

\subsection{Performance Diagram}
\label{A:PD}

The performance diagram, introduced in~\cite{kremenetski2021}, helps identify optimal values of \(\Delta_{\gamma, \beta}\) for LR-QAOA on a given problem instance.  
It visualizes how the approximation ratio varies as a function of the LR-QAOA depth \(p\) and the parameter scale \(\Delta_{\gamma, \beta}\).

\begin{figure}[!tbh]
\centering
\includegraphics[width=16cm]{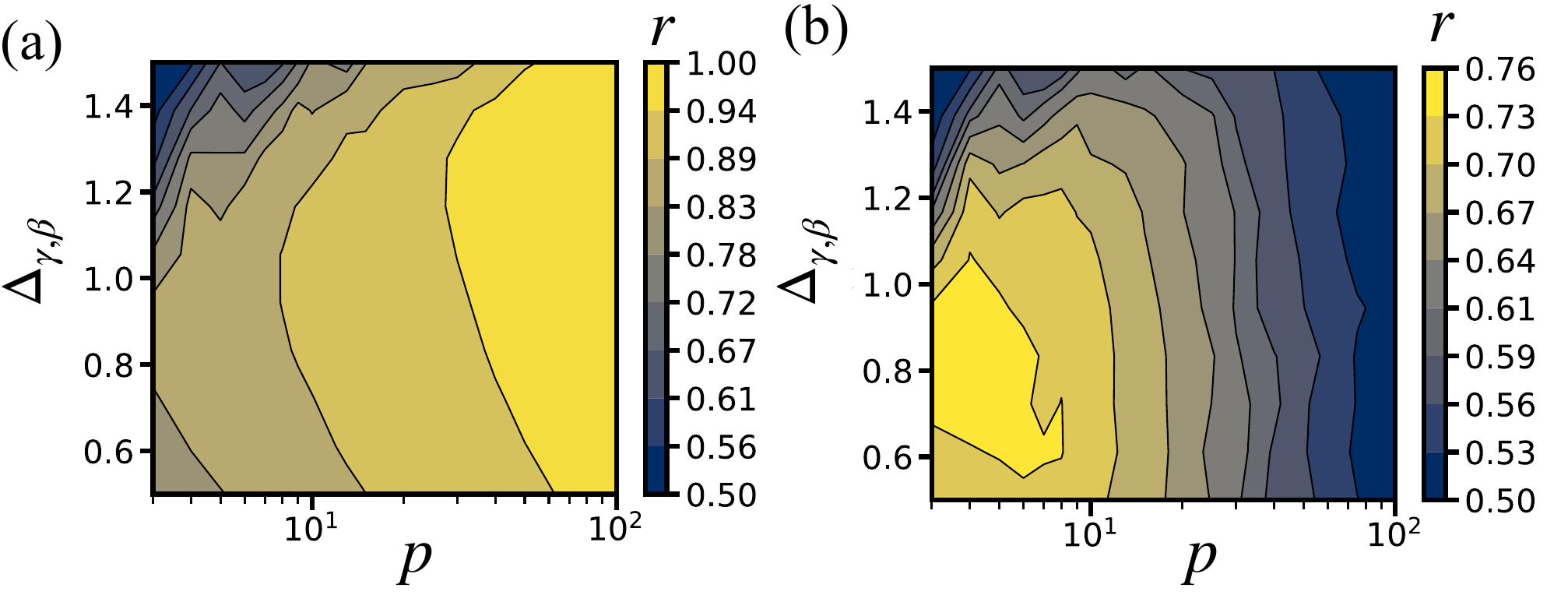}
\caption{\label{Fig:PD} 
Performance diagram for a 30-qubit 1D-chain WMC problem solved using LR-QAOA with \(p = 3\) to \(p = 100\), showing approximation ratio versus depth and \(\Delta_{\gamma,\beta}\).  
(a) State vector simulation (ideal case).  
(b) Real QPU: \texttt{ibm\_brisbane}.}
\end{figure}

Figure~\ref{Fig:PD}(a) shows the ideal case obtained from a noiseless state vector simulation.  
In the absence of noise, the approximation ratio improves monotonically with depth, reflecting coherent accumulation of algorithmic signal.  
In contrast, Fig.~\ref{Fig:PD}(b) shows the same diagram generated on a real QPU (\texttt{ibm\_brisbane}).  
Here, noise causes the approximation ratio to plateau or decline beyond a certain depth, and performance becomes sensitive to \(\Delta_{\gamma, \beta}\).  

On noisy devices, there is typically an optimal range of \(p\) values that yield the best approximation ratio.  
In this case, values between \(p = 3\) and \(p = 8\) perform similarly, but choosing the smallest such depth minimizes execution time and gate count while preserving solution quality.  
Performance diagrams are thus a valuable diagnostic tool for identifying useful depth regimes and parameter schedules on real QPUs.

\end{document}